\documentclass[a4paper,fleqn,usenatbib]{mnras}

\usepackage[T1]{fontenc}
\usepackage{ae,aecompl}

\usepackage{graphicx}	
\usepackage{amsmath}	
\usepackage{amssymb}	
\usepackage{subfig}
\usepackage{rotating}
\usepackage{pdflscape}

\title[Non-LTE Equivalent Widths for N\,{\sc ii}]{Non-LTE Equivalent Widths for N\,{\sc ii} with Error Estimates}

\author[A. Ahmed and T. A. A. Sigut]{A. Ahmed$^{1,2}$\thanks{E-mail:
ahamza5@uwo.ca (AA); asigut@uwo.ca (TAAS)} and T. A. A. Sigut$^{1,3}$ \\
$^{1}$Department of Physics and Astronomy, The University of Western Ontario \\ 
London, Ontario, N6A 3K7, Canada\\ 
$^{2}$Department of Astronomy, Cairo University, Giza, 12613, Egypt \\
$^{3}$Centre for Planetary Science and Exploration, The University of Western
Ontario \\ London, Ontario, N6A 3K7, Canada}

\date{Accepted 2015 October 7. Received 2015 October 7; in original form 2015 July 9}

\pubyear{2015}

\begin{document}
\label{firstpage}
\pagerange{\pageref{firstpage}--\pageref{lastpage}}
\maketitle

\begin{abstract}

Non-LTE calculations are performed for N\,{\sc ii} in stellar atmospheric
models appropriate to main sequence B-stars to produce new grids of
equivalent widths for the strongest N\,{\sc ii} lines commonly used
for abundance analysis.  There is reasonable agreement between our
calculations and previous results, although we find weaker non-LTE effects
in the strongest optical N\,{\sc ii} transition, $\lambda\,3995$.  We also
present a detailed estimation of the uncertainties in the equivalent
widths due to inaccuracies in the atomic data via Monte Carlo simulation
and investigate the completeness of our model atom in terms of included
energy levels. Uncertainties in the basic N\,{\sc ii} atomic data limit
the accuracy of abundance determinations to $\approx\,\pm0.10\,$dex at
the peak of the N\,{\sc ii} optical spectrum near $T_{\rm eff}\approx
24,000\,$K.

\end{abstract}

\begin{keywords}
stars: abundances - stars: atmospheres - radiative transfer -line: formation
\end{keywords}



\section{Introduction}

Accurate abundance measurements in the atmospheres of massive, upper
main sequence stars represent an important test of current models
of stellar evolution \citep{MM12,Pal13}.  Accurate CNO abundances,
and nitrogen abundances in particular, are of special significance
due to their potential role as diagnostics of rotational mixing
(e.g. \citet{heg00b}, \citet{heg00a}, \citet{mey00}, \citet{brott11},
\citet{Eks12}, \citet{GH14}, \citet{Mea14}). Stellar rotation
velocities reach their peak on the main sequence among the early-B and
late-O-type stars \citep{Fuk82}, and rapid rotation is predicted to mix
CNO-processed material into the stellar atmospheres, with the resulting
nitrogen enhancement being the easiest to detect \citep{tal97, mey00,
heg00a,brott11}. Searches for enhanced nitrogen abundances among main
sequence B stars have produced mixed results \citep{mae09,hun09,NP14},
and surprisingly, no nitrogen enhancements have been found among the
Be stars, which are the most rapidly-rotating population on the main
sequence \citep{len05,dun11}.

Accurate abundance analysis for B stars is complicated by several factors:
departures from local thermodynamic equilibrium (LTE), rapid rotation
(which introduces several issues-- see below), and, in cases such
as the Be stars, potential emission from circumstellar material.

Departure from LTE is a well known problem in early-type and evolved
stars \citep{mih78,kur79}.  In the non-LTE case, the calculation must
account for the non-local radiation field in the photosphere due to
photons coming from hotter, deeper layers and from photon loss through
the outer boundary. As a result, the level populations will differ from
the Saha-Boltzmann predictions at the local electron temperature and
density. To obtain the line source functions and optical depth scales
required for the line transfer problem, the coupled equations of radiative
transfer and statistical equilibrium must be solved in a self-consistent
manner \citep{mih78,can85}.

The large broadening of spectral lines in early-type stars due to rapid
rotation results in shallow and wide lines with low continuum contrast
and, potentially, strong line blending. In such cases, only a few, strong,
spectral lines of each element can be reliably measured.  The use of
only the strongest lines of a given element can compromise the accuracy
of the abundance analysis due to (typically) stronger non-LTE effects,
the confounding influence of microturbulence, and the dependence of
the equivalent widths on uncertain damping parameters. In addition,
the traditional method of estimating uncertainties from the dispersion
of the measured elemental abundance from many measured weak and strong
lines cannot be applied. In the case where only a few strong lines of
a given element are available, a detailed theoretical error analysis is
required to give the measured abundances meaning. This can be handled by
Monte Carlo simulation of the errors introduced by uncertain atomic data
(including damping widths), uncertain stellar parameters ($T_{\rm eff}$,
$\log\,g$, and the microturbulence), and the uncertainty in the measured
equivalent width \citep{sig96}.

Complications due to rapid rotation include gravitational
darkening, where the stellar surface distorts and the temperature and
gravity become dependent on latitude \citep{von24}. As a consequence, the
strengths of spectral lines will be dependent on the stellar inclination
\citep{sto87}. Finally, in the case of the Be stars, there is the
possibility of contamination by circumstellar material \citep{por03}.

In this paper, the focus is on the limiting accuracy of predicted N\,{\sc
ii} equivalent widths due to uncertainties in the basic atomic data used
in the non-LTE calculation. The methodology follows that of \citet{sig96},
and the structure of the paper is as follows: In Section~(\ref{sec:previous}),
we give a brief summary of previous non-LTE calculations for N\,{\sc
ii}. In Section~(\ref{N-atom}), we discuss the atomic data used to
construct our nitrogen atom. In Section (\ref{result}), the results of
our non-LTE nitrogen calculations are given, and the error bounds on the
predicted equivalent widths due to random errors and systematic errors
are discussed. Section (\ref{disc}) gives conclusions.

\section{Previous Work}
\label{sec:previous}

\citet{duf79} investigated the atmospheric nitrogen abundance for a
number of main-sequence B-type stars using the complete linearization
method \citep{auer73} and a 13 level N\,{\sc ii} atom, which included
only singlet states.  \citet{duf81} extended this work by including the
additional 14 lowest energy triplet levels. The calculation included 44
allowed radiative transitions, with the rates for 9 fully linearized.
Non-LTE and LTE equivalent widths for three singlet lines ($\lambda\,3995$, 
$\lambda\,4447$, and $\lambda\,4228$)\footnote{All wavelengths in this paper
are given in Angstroms, unless otherwise noted.} and three triplet lines (
$\lambda\,4631$, $\lambda\,5045$ and $\lambda\,5680$)
were calculated for stellar $T_{\rm eff}$ between 20,000 and 32,500 K.
In general, the predicted non-LTE equivalent widths were significantly
stronger than the corresponding LTE values and the 
difference increased with $T_{\rm eff}$.

An extensive, nitrogen atom was introduced by \citet{bb88}
and \citet{bb89}.  They constructed non-LTE and LTE equivalent width
grids of 35 N\,{\sc ii} radiative transitions in the wavelength region
between 4000\,\AA\;and 5000\,\AA, calculated over stellar $T_{\rm eff}$
between 24,000 and 33,000 K.  In this work, the non-LTE populations of
energy levels with principle quantum number up to 4 were included for
N\,{\sc i} and N\,{\sc ii}, and the lowest five levels of N\,{\sc iii}
and the ground level of N\,{\sc iv} were included in the linearization
method.  Results showed that there was strong non-LTE strengthening
of the predicted equivalent widths for some lines that could lower the
estimated nitrogen abundance by up to 0.25 dex.

\citet{kor99} investigated the nitrogen abundance of $\gamma\,$Peg
(B2{\sc V}) in order to test nitrogen enrichment due to rotational
mixing, For this purpose, a nitrogen atom was constructed which
consisted of 109 levels: 3 ground levels of N\,{\sc i}, the 93 lowest
energy levels of N\,{\sc ii}, the 12 lowest levels of N\,{\sc iii},
and the ground state of N\,{\sc iv}.  This calculation included all
allowed radiative transitions with wavelengths less than 10 $\rm \mu m$,
with 92 transitions computed in detail; the rates for the rest were
kept fixed.  \citet{kor99} also provided non-LTE and LTE equivalent grids
for 23 N\,{\sc ii} transitions at stellar $T_{\rm eff}$ between 16,000
and 32000\,K.  In general, there was good agreement with \citet{bb88},
but with some differences: firstly, the difference between LTE and
non-LTE equivalent widths were larger than those of \citet{bb88},
and the maximum differences occurred at lower $T_{\rm eff}$; secondly,
the maximum calculated equivalent widths occurred at lower $T_{\rm eff}$,
and this was attributed to the different (fixed, LTE) model atmospheres
used in the works.

Finally, \citet{prz01} performed non-LTE line formation for N\,{\sc
i}/N\,{\sc ii} in order to determine the nitrogen abundance of a number of
A and B type stars: Vega (A0V), and four late-A and early-B supergiants. In
this work, an extensive nitrogen atom was used with recent and accurate
atomic data. This work was mainly focused on studying objects with low
temperatures, $\rm T_{eff}\, \leq \,12,000$, where N\,{\sc ii} is not
the dominant ionization stage. They found weak non-LTE effects on
the N\,{\sc ii} lines and suggested further investigation at higher
effective temperatures.

In conclusion, many previous studies have investigated the non-LTE problem
of N\,{\sc ii}, aiming to get accurate equivalent widths using improved
techniques and more accurate atomic data. However, none of these works
present a detailed analysis of the uncertainties of the estimated
equivalent widths which is the main objective of the current work.

\section{The Nitrogen atom}
\label{N-atom}

\subsection{N\,{\sc ii} Atomic data}

\begin{table}
\centering{
\caption{Energy level data for the lowest 16 LS states of N\,{\sc ii} and
the ground states of N\,{\sc iii} and N\,{\sc iv}.  \label{NII.Elevels}}}
\begin{center}
{\small
\begin{tabular}{crrrl}
\hline\hline
n & \multicolumn{1}{c}{Energy $(\rm cm^{-1})$} & g & $\lambda_{\rm thres}$(\AA)    & Configuration\\
\hline
1 & $0.000$         &  $9.0$ & \multicolumn{1}{l}{\;\;418.8}     & \multicolumn{1}{l}{$\;2p^2\;\;^3P\;\; $} N\,{\sc ii}     \\
2 & $15316.200$     & $5.0$     & \multicolumn{1}{l}{\;\;447.6}     & \multicolumn{1}{l}{$\;2p^2\;\;^1D\;\;\;\;$}\\
3 & $32688.801$     & $1.0$     & \multicolumn{1}{l}{\;\;485.3}     & \multicolumn{1}{l}{$\;2p^2\;\;^1S\;\;\;\;$}\\
4 & $ 46784.602$ &  $  5.0$ & \multicolumn{1}{l}{\;\;520.9} & \multicolumn{1}{l}{$\;2p^3\;\;^5S^o\;\;$} \\ 
5 & $ 92244.484$ &  $ 15.0$ & \multicolumn{1}{l}{\;\;682.6} & \multicolumn{1}{l}{$\;2p^3\;\;^3D^o\;\;$} \\
6 & $109217.922$ &  $  9.0$ & \multicolumn{1}{l}{\;\;772.0} & \multicolumn{1}{l}{$\;2p^3\;\;^3P^o\;\;$}\\
7 & $144187.938$     & $5.0$    & \multicolumn{1}{l}{1057.5}     & \multicolumn{1}{l}{$\;2p^3\;\;^1D^o\;\;\;$}\\
8 & $149012.406$     & $9.0$     & \multicolumn{1}{l}{1114.4}    & \multicolumn{1}{l}{$\;3s\;\;\;^3P^o\;\;$} \\
9 & $149187.797$     & $3.0$     & \multicolumn{1}{l}{1116.5}    & \multicolumn{1}{l}{$\;3s\;\;^1P^o\;\;$} \\
10 & $155126.734$ & $3.0$     & \multicolumn{1}{l}{1195.8}    & \multicolumn{1}{l}{$\;2p^3\;\;^3S^o\;\;$}\\
11 & $164610.766$ & $3.0$     & \multicolumn{1}{l}{1348.8}    & \multicolumn{1}{l}{$\;3p\;\;\;^1P\;\;\;$} \\ 
12 & $166615.188$ & $15.0$ & \multicolumn{1}{l}{1386.3}    & \multicolumn{1}{l}{$\;3p\;\;^3D\;\;$}     \\
13 & $166765.656$ & $3.0$     & \multicolumn{1}{l}{1389.2}    & \multicolumn{1}{l}{$\;2p^3\;\;^1P^o\;\;$}\\
14 & $168892.203$ & $3.0$     & \multicolumn{1}{l}{1431.5}    & \multicolumn{1}{l}{$\;3p\;\;\;^3S\;\;$}     \\
15 & $170636.375$ & $9.0$     & \multicolumn{1}{l}{1468.1}    & \multicolumn{1}{l}{$\;3p\;\;\;^3P\;\;$}     \\
16 & $174212.031$ & $5.0$     & \multicolumn{1}{l}{1549.5}     & \multicolumn{1}{l}{$\;3p\;\;^1D\;\;\;$} \\
\ldots   &              &           &                                &                                       \\
94 & $238750.300$ & $15.0$ & \multicolumn{1}{l}{\;\;261.3}    & \multicolumn{1}{l}{$\;2p\;\;\;^2P^o\;\;$} N\,{\sc iii} \\
\ldots   &              &           &                                &                                       \\
106 & $621454.625$ & $1.0$ & \multicolumn{1}{l}{\;\;160.0}    & \multicolumn{1}{l}{$\;2s^2\;\;^1S\;\;$} N\,{\sc iv} \\
\hline
\end{tabular}}
\end{center}
\vspace{0.1in}
\end{table}

Table~\ref{NII.Elevels} lists the experimental values for the first 16
N\,{\sc ii} energy levels, taken from \citet{moo93} and available through
NIST database.\footnote{http://www.nist.gov/pml/data/asd.cfm} The N\,{\sc ii} atom itself includes 93 energy levels,
complete through $n=6$.  The oscillator strengths and photoionization
cross-sections were taken from the Opacity Project \citep{luo89},
through the TOPBASE database \citep{cun93}. In total, 580 radiative
transitions were included in the calculation, representing all
transitions between the included energy levels with $f\,\geq\,10^{-3}$.
Table~\ref{rbb-trans-data} lists the atomic data for a number of
bound-bound radiative transitions commonly used in N\,{\sc ii} abundance
determinations.
Note that the non-LTE calculation computes populations
for the total LS energy levels; populations for fine structure levels are
obtained by assuming these levels are populated relative to their
statistical weights. This is a very good approximation in a stellar
atmosphere due to the small energy spacing of the fine structure levels
and the large rates of collisional transitions between these levels.

\begin{table}
\centering{
\caption{Atomic data for several fine-structure N\;{\sc ii} transitions of interest.
\label{rbb-trans-data}}}
\begin{center}
{\small
\begin{tabular}{rrlcr}
\hline\hline
$\lambda\,$(\AA) & $A_{ul} (s^{-1})$ & Transition & $(l-u)$ & $\gamma_s$ (\AA) \\
\hline
3955.0 &   $1.203^{+07}$& $3s \,^3P^o\,(1)\rightarrow\,3p \,^1D\,(2)$& 8 - 16 &$3.9^{-02}$\\
3995.0 &   $1.386^{+08}$ & $3s \,^1P^o\,(1)\rightarrow\,3p \,^1D\,(2)$&  9 - 16 &$3.9^{-02}$\\
4601.5 &   $2.325^{+07}$ & $3s \,^3P^o\,(1)\rightarrow\,3p \,^3P\,(2)$&  8 - 15 &$4.5^{-02}$\\
4607.2 &  $3.310^{+07}$ &\multicolumn{1}{l}{$\hspace{0.9cm}(0)\rightarrow\hspace{0.9cm}(1)$} & &\\
4613.9 &  $2.227^{+07}$ &\multicolumn{1}{l}{$\hspace{0.9cm}(1)\rightarrow\hspace{0.9cm}(1)$} & &\\
4621.4 &  $9.474^{+07}$ &\multicolumn{1}{l}{$\hspace{0.9cm}(1)\rightarrow\hspace{0.9cm}(0)$} & &\\
4630.5 &  $7.878^{+07}$ &\multicolumn{1}{l}{$\hspace{0.9cm}(2)\rightarrow\hspace{0.9cm}(2)$} & &\\
4643.1 &  $4.611^{+07}$ &\multicolumn{1}{l}{$\hspace{0.9cm}(2)\rightarrow\hspace{0.9cm}(1)$} & &\\
4447.0 &   $1.174^{+08}$ &$3p \,^1P\;\;(1)\rightarrow\,3d \,^1D^o(2)$& 11 - 19 &$9.0^{-02}$\\
4987.4 &  $7.474^{+07}$ &$3p \,^3S\;\;(1)\rightarrow\,3d \,^3P^o\,(0)$& 14 - 21 &$7.0^{-02}$\\
4994.4 &   $7.583^{+07}$ &\multicolumn{1}{l}{$\hspace{0.9cm}(1)\,\rightarrow\hspace{0.9cm}(1)$} & &\\
5007.3 &   $7.956^{+07}$ &\multicolumn{1}{l}{$\hspace{0.9cm}(1)\,\rightarrow\hspace{0.9cm}(2)$} & &\\
5001.1 &  $9.719^{+07}$ &$3p \,^3D\;\;(1)\rightarrow\,3d \,^3F^o\,(2)$& 12 - 18 &$6.7^{-02}$\\
5001.5 &  $1.046^{+08}$ &\multicolumn{1}{l}{$\hspace{0.9cm}(2)\,\rightarrow\hspace{0.9cm}(3)$} & &\\
5005.1 &  $1.155^{+08}$ &\multicolumn{1}{l}{$\hspace{0.9cm}(3)\,\rightarrow\hspace{0.9cm}(4)$} & &\\
5016.4 &  $1.581^{+07}$ &\multicolumn{1}{l}{$\hspace{0.9cm}(2)\,\rightarrow\hspace{0.9cm}(2)$} & &\\
5025.7 &  $1.055^{+07}$ &\multicolumn{1}{l}{$\hspace{0.9cm}(3)\,\rightarrow\hspace{0.9cm}(3)$} & &\\
5040.7 &  $4.722^{+05}$ &\multicolumn{1}{l}{$\hspace{0.9cm}(3)\,\rightarrow\hspace{0.9cm}(2)$} & &\\
5002.7 &  $8.661^{+06}$ &$3s \,^3P^o\,(0)\rightarrow\,3p \,^3S\,\;(1)$& 8 - 14 &$5.0^{-02}$\\
5010.6 &  $2.165^{+07}$ &\multicolumn{1}{l}{$\hspace{0.9cm}(1)\,\rightarrow\hspace{0.9cm}(1)$} & &\\
5045.1 &  $3.481^{+07}$ &\multicolumn{1}{l}{$\hspace{0.9cm}(2)\,\rightarrow\hspace{0.9cm}(1)$} & &\\
5666.6 &  $3.608^{+07}$ &$3s \,^3P^o\,(1)\rightarrow\,3p \,^3D\,\;(2)$& 8 - 12 &$6.1^{-02}$\\
5676.0 &  $2.916^{+07}$ &\multicolumn{1}{l}{$\hspace{0.9cm}(0)\,\rightarrow\hspace{0.9cm}(1)$} & &\\
5679.6 &  $5.194^{+07}$ &\multicolumn{1}{l}{$\hspace{0.9cm}(2)\,\rightarrow\hspace{0.9cm}(3)$} & &\\
5686.2 &  $1.875^{+07}$ &\multicolumn{1}{l}{$\hspace{0.9cm}(1)\,\rightarrow\hspace{0.9cm}(1)$} & &\\
5710.8 &  $1.229^{+07}$ &\multicolumn{1}{l}{$\hspace{0.9cm}(2)\,\rightarrow\hspace{0.9cm}(2)$} & &\\
5730.7 &  $1.221^{+06}$ &\multicolumn{1}{l}{$\hspace{0.9cm}(2)\,\rightarrow\hspace{0.9cm}(1)$} & &\\
6482.1 &  $2.913^{+07}$ &$3s \,^1P^o\,(1)\rightarrow\,3p \,^1P\;\;(1)$& 9 - 11 &$1.3^{-01}$\\
\hline
\end{tabular}}
\end{center}
\vspace{0.1in}
\flushleft
\noindent{Note: Stark widths, $\gamma_s$, were calculated assuming an electron number density of $10^{+16}\;\rm cm^{-3}$ and a temperature of 30,000~K. The J values of the fine structure levels of each LS state are shown in brackets.}
\end{table}

Thermally-averaged collision strengths for bound-bound collisional
transitions between the lowest 23 LS states of N\,{\sc ii}  were
taken from \citet{hud04,hud05a}. These were calculated in a 23 state,
close-coupling calculation using the ${\cal R}$-Matrix method. The impact
parameter approximation of \citet{sea62} was used for the remaining
bound-bound collision strengths for allowed transitions. The collision
strengths for forbidden transitions were assumed to be 0.1. The rates
of collisional ionization of N\,{\sc ii} energy levels to the  N\,{\sc
iii} ground state were estimated using the procedure of \citet{sea62},
where the rate is proportional to the photoionization cross section at
threshold, as given in \citet{Jef68}.

\begin{figure}
\includegraphics[scale= 0.45]{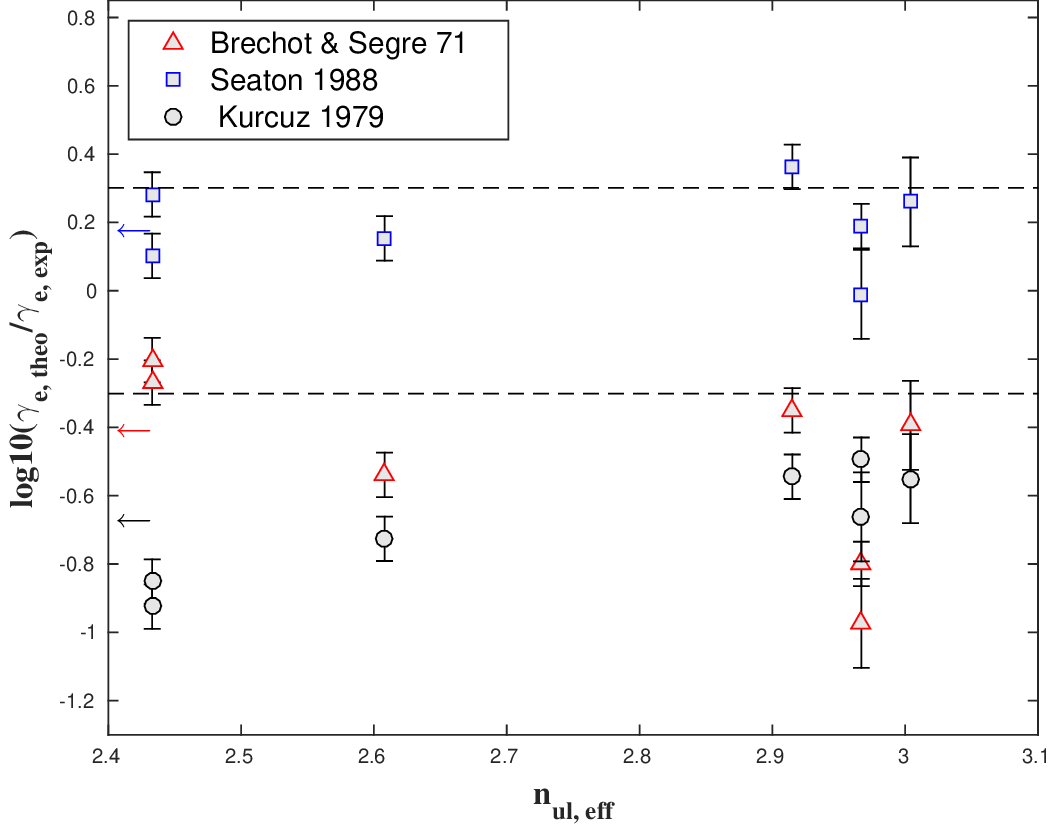}
\caption{Ratio of theoretical-to-experimental quadratic Stark width versus
the effective quantum number of the upper level at $T_{\rm e}=28\,000\;$K. 
The theoretical
widths were calculated using three different approximations with the
sources noted in the legend. The dotted lines indicate a factor of two uncertainty.
\label{gamma_ratio_comp}}
\end{figure}

The line profiles for all radiative transitions included natural
broadening, thermal broadening, and pressure broadening due to collisions
with charged and neutral particles.  Quadratic Stark broadening,
due to quasi-static collisions with electrons, represents the most
important collisional contributor to the line width in the atmospheres
of hot stars. As the N\,{\sc ii} transitions are often strong, it
is important to have accurate Stark widths. For this reason, we have
calculated the Stark widths using the method developed for the Opacity
Project \citep{sea88}. We compare these Stark widths to the experiment
values of \cite{kon02} and to calculations using the semi-classical
approximation of \citet{SBS71} and the commonly-used formula of
\citet{kur79}, that represents a fit of the widths of \citet{SBS71},
in Figure~\ref{gamma_ratio_comp}.  The comparison assumes $T_{\rm e}$=
28,000 K and the figure shows Stark width versus the effective principle
quantum number ($n^{\rm eff}$) of the upper level, defined for the
$i^{th}$ energy level as
\begin{equation*} 
(n^{\rm eff}_i)^2=\frac{Z\,{\cal R}}{(I-E_i )}\,, 
\end{equation*} 
where $\cal R$ is the Rydberg constant, $I$ and $E_i$ are the ionization
energy of the atom and the energy of the $i^{th}$ state, and $Z$ is the
core charge.
Figure~\ref{gamma_ratio_comp} shows that the Stark widths
calculated following \citet{sea88} agree best with the experimental
values of \citet{kon02} at $T_{\rm e}=$ 28\,000 K. However, at
lower temperatures, the \citet{sea88} and \citet{SBS71} methods give
similar accuracies compared to experimental values, with the
\citet{sea88} formula tending to be higher than experiment and the
\citet{SBS71} method. We note that the formula adopted by
\citet{kur79} as an overall fit to the \citet{SBS71} gives
Stark widths that seem too small.

Finally, the thermal widths of the lines included the contribution
of microturbulence, $\xi_t$. Microturbulence is an important physical
process in stellar atmospheres that can affect the strength of stronger
spectral lines \citep{gra92}. Microturbulence represents the dispersion
of a non-thermal velocity field on a scale smaller than unit optical
depth which acts to broaden the atomic absorption. Microturbulence
is a confounding factor in abundance determinations. As weak lines
on the linear portion of their curves of growth are insensitive to
microturbulence, forcing the abundances from strong and weak lines to
agree can fix its value.  However, if only strong lines are available
for the abundance analysis, microturbulence may represent a serious
limitation to the achievable accuracy. As to its physical origin,
\citet{cant09} show that the iron-peak in stellar opacities can lead to
the formation of convective cells close to the surface of hot stars which
could be the origin of such turbulence, through energy dissipated
from gravitational and pressure waves propagating outwardly from these
convective cells.

%
%

%
%
\begin{table}
\centering{
\caption{Average ratios of calculated to experimental Stark widths.
S88 refers to \citet{sea88}, K79 refers to
\citet{kur79}, and SB71 refers to \citet{SBS71}.
\label{gamma_ratios}}}
\begin{center}
{\small
\begin{tabular}{lrrr}
\hline\hline
Temperature  & \multicolumn{3}{c}{Average $\left(\gamma_{\rm calc}/\gamma_{\rm exp}\right)$} \\
(K) & S88 & K79 & SB71\\
\hline 
8000 	& 2.12 & 0.23 & 0.71\\
15000 & 1.92 &	0.22 & 0.61\\ 
28000 & 1.55 & 0.21 & 0.40\\
\hline
\end{tabular}}
\end{center}
\vspace{0.1in}
\end{table}

\subsection{N\,{\sc iii} \& N\,{\sc iv} Atomic Data}

N\,{\sc iii} and N\,{\sc iv} energy levels were taken from
\citet{moo93}. In total, the twelve lowest LS levels of N\,{\sc iii}
and the ground state of N\,{\sc iv} were included in the calculation.
The oscillator strengths of radiative bound-bound transitions of N\,{\sc
iii} were obtained from \citet{bell95} and \citet{fer99}, available from
the NIST database. The photoionization cross-sections were also taken
from \citet{fer99}. Similarly, the
collision strengths of excitation and ionization were calculated using
impact parameter approximation of \citet{sea62}.

\section{Calculations}\label{result}

The N\,{\sc ii} non-LTE line formation calculation was carried out for
all combinations of nine $T_{\rm eff}$, between 15,000 and 31,000 K in
steps of 2000~K, three surface gravities, $\log g$ (3.5, 4 and 4.5), four
microturbulent velocities, $\xi_t$ (0, 2, 5 and $10\;\rm km\,s^{-1}$),
and 7 nitrogen abundances, between 6.83 and 8.13~dex. The MULTI code,
v2 \citep{car92}, was used. MULTI solves the statistical equilibrium
and radiative transfer equations simultaneously in an iterative method
using the approximate lambda-operator technique of \cite{sch81}. The
option of using a local approximate lambda-operator was used in the
current calculations.  A solar nitrogen abundance $\epsilon_{N}$= 7.83
was adopted from \cite{gre10}.

The fixed, background model photospheres providing $T(\tau)$ and $P(\tau)$
for the non-LTE calculation were taken from the LTE, line blanketed,
atmospheres of ATLAS9 \citep{kur93}. ATLAS9 was also used to provide the
mean intensity within the stellar atmosphere as a function of optical
depth, $J_{\nu}(\tau)$, used to compute all of the photoionization and
recombination rates which were kept fixed during the calculations.
Using LTE model atmospheres can be a source of error, particularly for
hotter stars; however, previous studies have shown that a comprehensive
inclusion of line-blanketing is more important than non-LTE effects up
to stellar effective temperatures of $\approx\,$30,000 K \citep{prz01}.

As MULTI was originally developed for the atmospheres of cool stars,
modification to the background opacities are required in order to
be suitable for early-type stars; its default opacity package was
replaced with the extensive package that is available through ATLAS9
\citep{sig96b}.

\subsection{Ionization Balance and Departure Coefficients}

\begin{figure}
\centering
\subfloat[]{
\includegraphics[scale= 0.45]{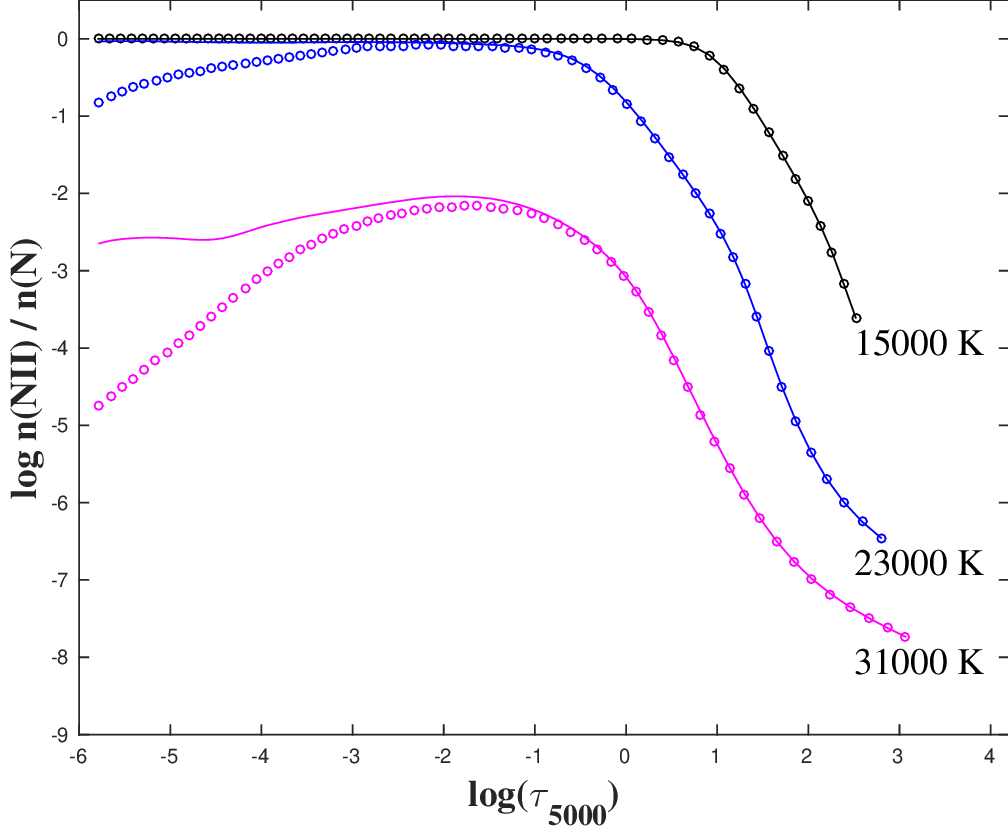}
\label{N2_frac}
}
\vspace{0.15cm}
\subfloat[]{
\includegraphics[scale= 0.45]{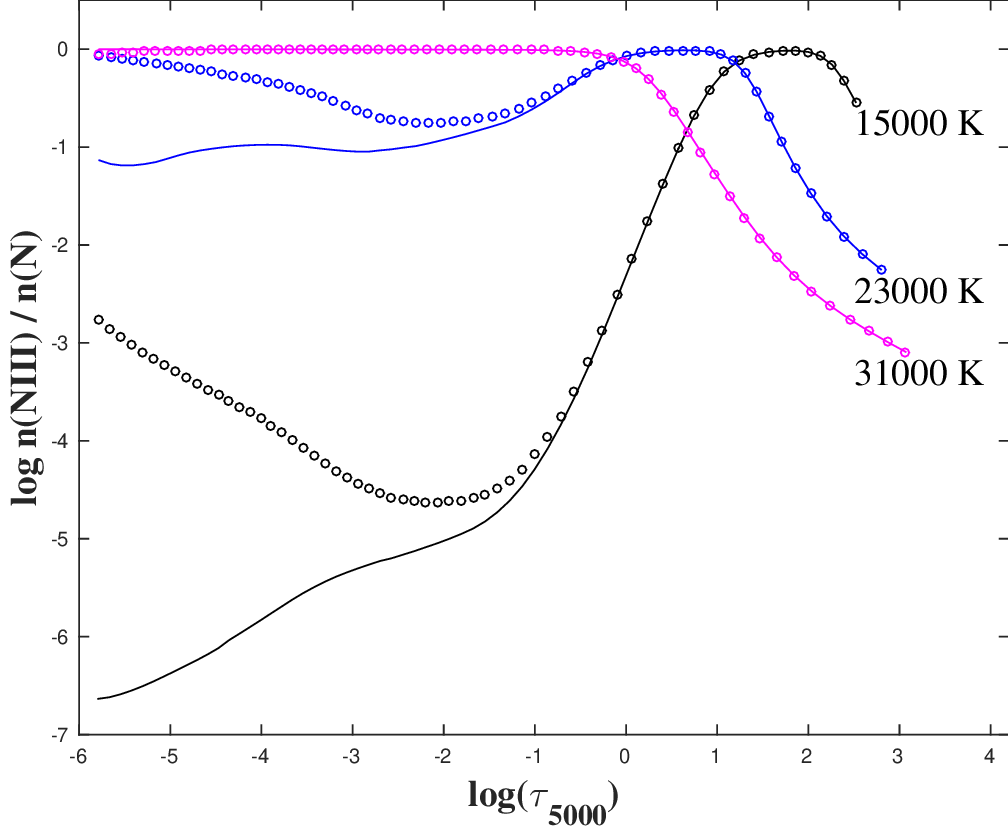}
\label{N3_frac}
}
\caption{Panel~(a) shows the fraction of N\,{\sc ii}
as a function of $\log\tau_{5000}$ for several stellar $T_{\rm eff}$
in atmospheres with $\log g$= 4.0, $\xi_t= 5\;\rm km\,s^{-1}$, and
$\epsilon_N$= 7.83. The circles represent the non-LTE fraction and
solid lines, the LTE fractions. Panel~(b) is the same,
but for N\,{\sc iii}. In both panels, the line colour indicates $T_{\rm eff}$ 
as 15000\,K (black), 23,000\,K (blue), and 31,000\,K (purple).\label{ion_frac}}
\end{figure}

Figure~\ref{ion_frac} shows the predicted LTE and non-LTE ionization
fractions of N\,{\sc ii} (top panel) and N\,{\sc iii} (bottom panel) as a
function of the continuum optical depth at $5000\,$\AA, $\log\tau_{5000}$,
for the range of the $T_{\rm eff}$ considered. These illustrated models
assume $\log\,g=4.0$, $\xi_t= 5\;\rm km\,s^{-1}$, and the solar nitrogen
abundance.

At $T_{\rm eff}$ less than $\sim 21,000\,$K, N\,{\sc ii} is the dominant
ionization stage throughout the formation region of the optical N\,{\sc
ii} lines, $-2\le\,\log \tau_{5000} \le 0$, and there is little deviation
from the LTE ionization fraction. Table~\ref{NII.Elevels} shows that the
six lowest energy levels of N\,{\sc ii} have photoionization thresholds
shortward of the Lyman limit at 912\,\AA, which remains optically
thick throughout most of the atmosphere. This leads to a strongly local
photoionizing radiation field, i.e.\ $J_{\nu}\,\approx\,B_{\nu}(T_e)$.
It is only by level 7, $2p\,3p\,^1D$, that the photoionization threshold
begin to lie in the short-wavelength region of the Balmer continuum where
the photoionizing radiation field can be both hot and non-local. This high
radiation temperature in the Balmer continuum can drive overionization;
however, even at the highest $T_{\rm eff}$ considered here, the predicted
non-LTE over-ionization of N\,{\sc ii} is quite small in the line
formation region (Figure~\ref{ion_frac}).

\begin{figure*}
\centering
\subfloat[]{
\includegraphics[scale= 0.4]{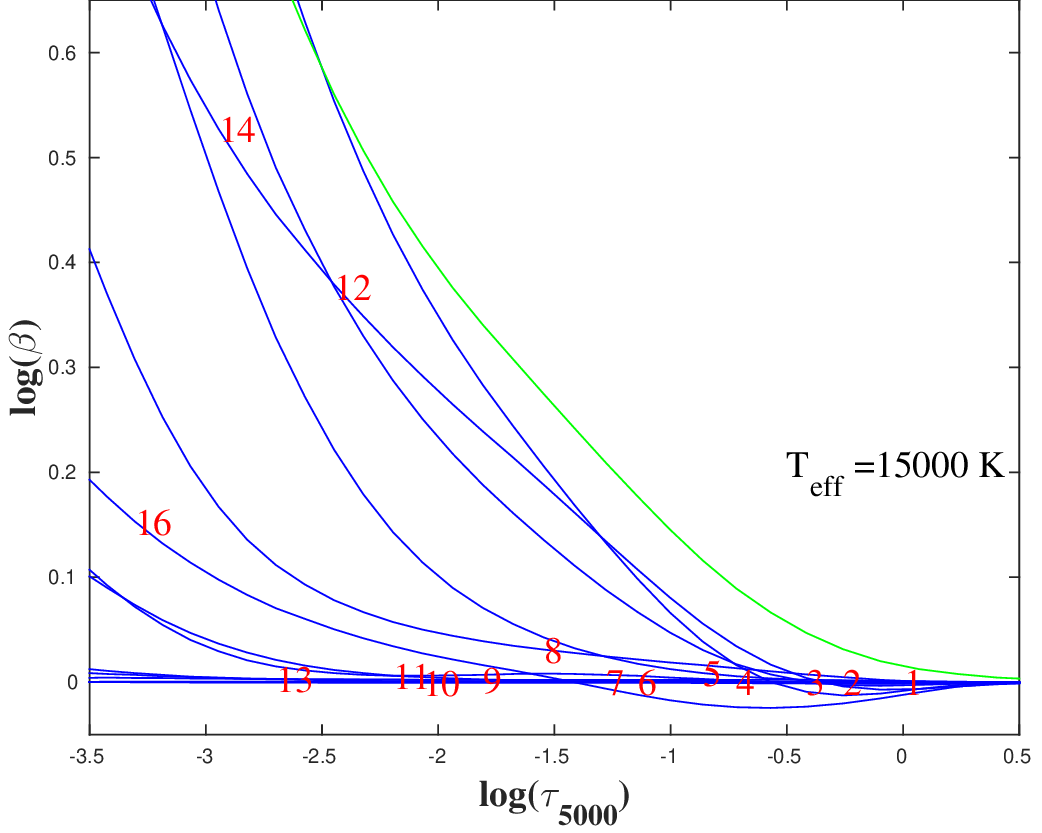}
\label{Departure_Coeff_T15000}
}
\hspace{0.1cm} 
\vspace{0.005cm}
\subfloat[]{
\includegraphics[scale= 0.4]{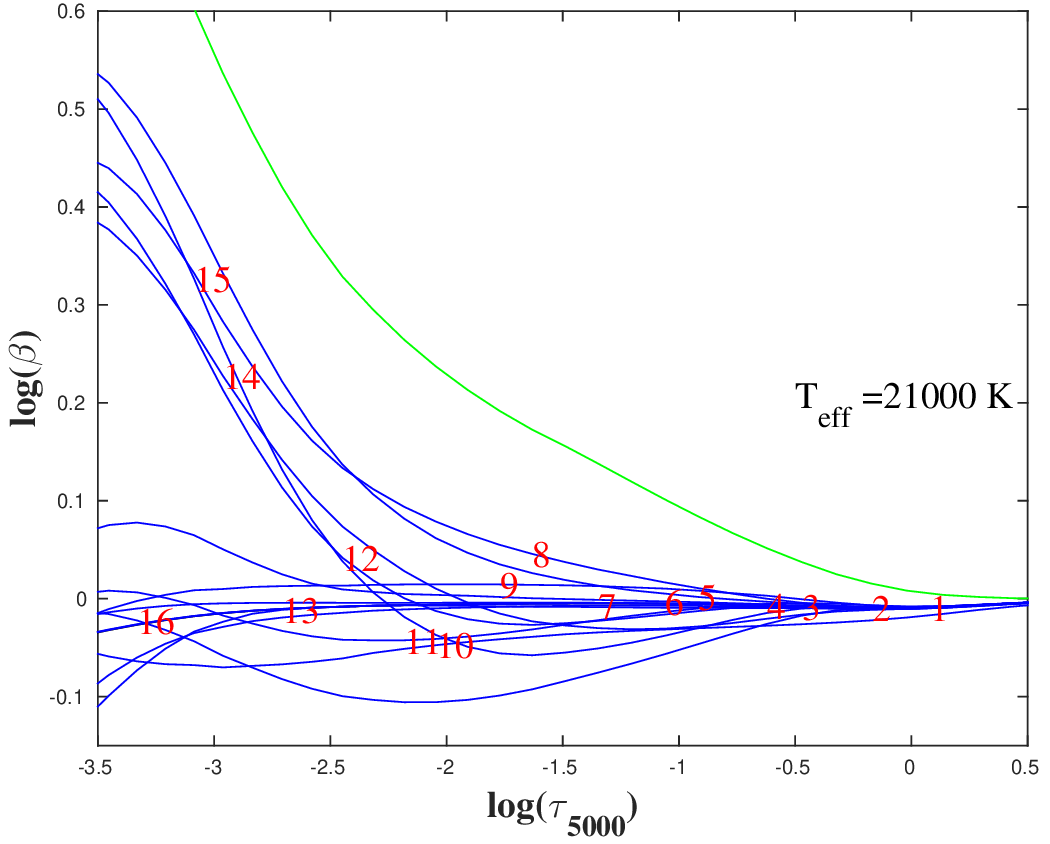}
\label{Departure_Coeff_T21000}
}
\vspace{0.005cm}
\subfloat[]{
\includegraphics[scale= 0.4]{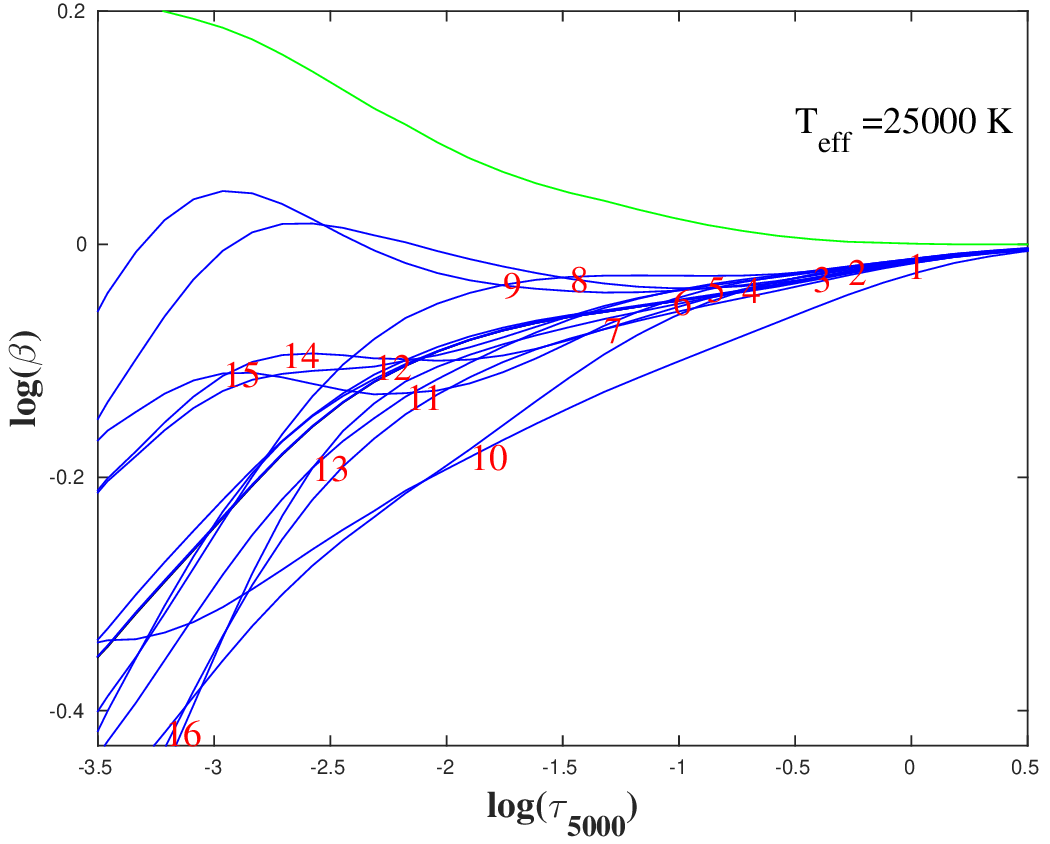}
\label{Departure_Coeff_T25000}
}
\hspace{0.1cm}
\subfloat[]{
\includegraphics[scale= 0.4]{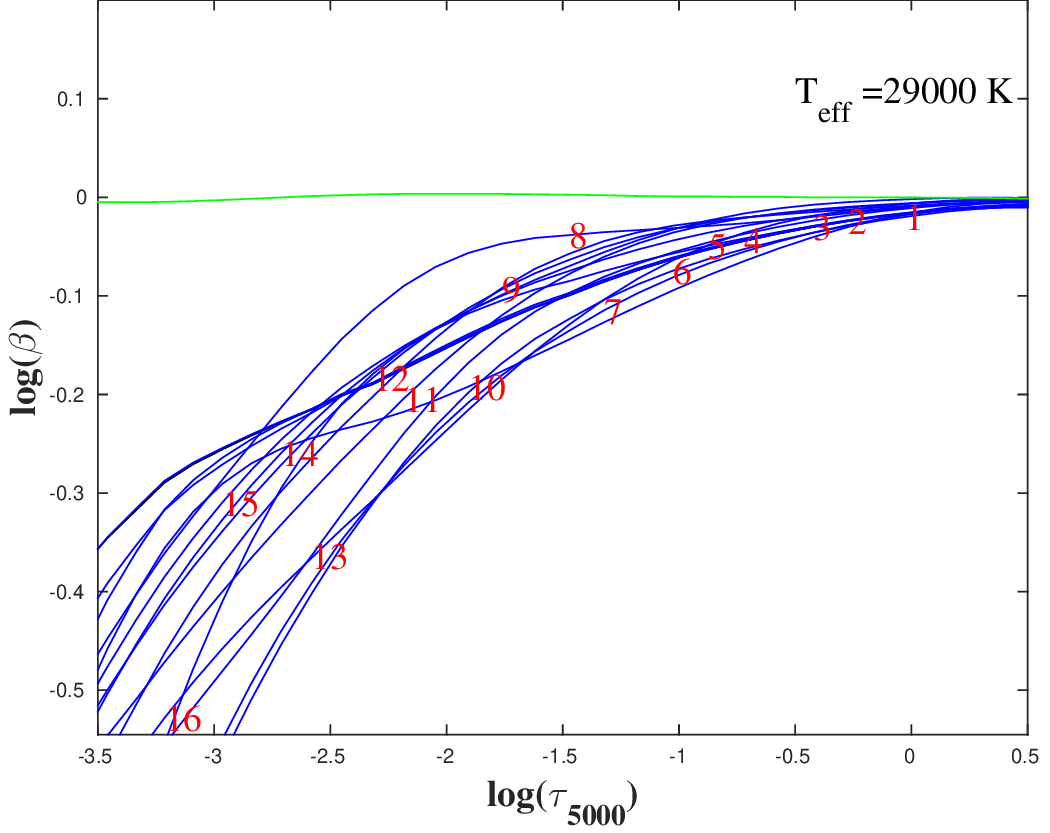}
\label{Departure_Coeff_T29000}
}
\caption{Non-LTE departures coefficients, $\log\beta_i$, for the
lowest 16 energy levels of N\,{\sc ii} of Table~\ref{NII.Elevels} (blue
lines identified by level number) and the ground level of N\,{\sc iii}
(green line) as a function of $\tau_{5000}$. The stellar $T_{\rm eff}$
shown are 15,000\,K (panel~a), 21,000\,K (panel~b), 25000\,K (panel~c),
and 29,000\,K (panel~d).  The surface gravity and microturbulent velocity
were fixed at $\log\,g=4.0$ and $\xi_t=5\;\rm km\,s^{-1}$, and the solar
nitrogen abundance was assumed. \label{Departure_Coeffs}}
\end{figure*}

The predicted departure coefficients for the 16 lowest LS states of N\,{\sc
ii} (Table~\ref{NII.Elevels}) and the ground state of N\,{\sc iii},
for four values of $T_{\rm eff}$
in models with $\log g=$ 4.0, $\xi_t= 5.0\;\rm km\,s^{-1}$, and the
solar nitrogen abundance, are shown in Figure \ref{Departure_Coeffs}. The
departure coefficient of the $i^{th}$ energy level, $\beta_i$, is defined
as the ratio of the non-LTE number density to the corresponding
LTE value computed by the Saha/Boltzmann equations for the local values
of $T_e$ and $N_e$,
\begin{equation} 
\beta_i\equiv\frac{n_i}{n_i^*(T_e,N_e)}\;, 
\end{equation} 
where $n_i^*$ and $n_i$ are the predicted LTE and non-LTE number densities
of the $i^{th}$ level, respectively.  Among the levels of particular
interest are 9 and 16, the upper and lower levels of $\lambda\,3995$,
and 9 and 11, the upper and lower levels of $\lambda\,6482$
(see Table~\ref{rbb-trans-data}). These transitions will be explicitly
discussed in the remainder of the text.

At $T_{\rm eff}=15,000\,$K, Figure~\ref{Departure_Coeff_T15000},
all of the low-lying levels ($\le 10$) are in LTE throughout the
atmosphere. As $T_{\rm eff}$ increases, there is a systematic trend
for overionization of all of the low-lying levels to set in. This is clear
by $T_{\rm eff}=25,000\,$K, Figure~\ref{Departure_Coeff_T25000}, where
near $\log\tau_{5000}\approx -0.5$, all of the low-lying levels share the
same departure coefficient, $\beta$, due to strong collisional coupling,
with $\beta<1$ due to photoionization in the short wavelength portion
of the Balmer continuum. By $T_{\rm eff}=25,000\,$K, the upper levels
with threshold in this region have sufficient population and N\,{\sc ii}
is no longer the dominant ionization stage, allowing the overionization
to occur. This trend continues for $T_{\rm eff}=29,000\,$K. For $T_{\rm
eff}<21,000\,$K, the overpopulation of the N\,{\sc iii} ground state leads
to general over population of the higher N\,{\sc ii} excited states,
although this effect diminishes by $T_{\rm eff}=25,000\,$K.

\begin{figure*}
\centering
\subfloat[]{
\includegraphics[scale= 0.4]{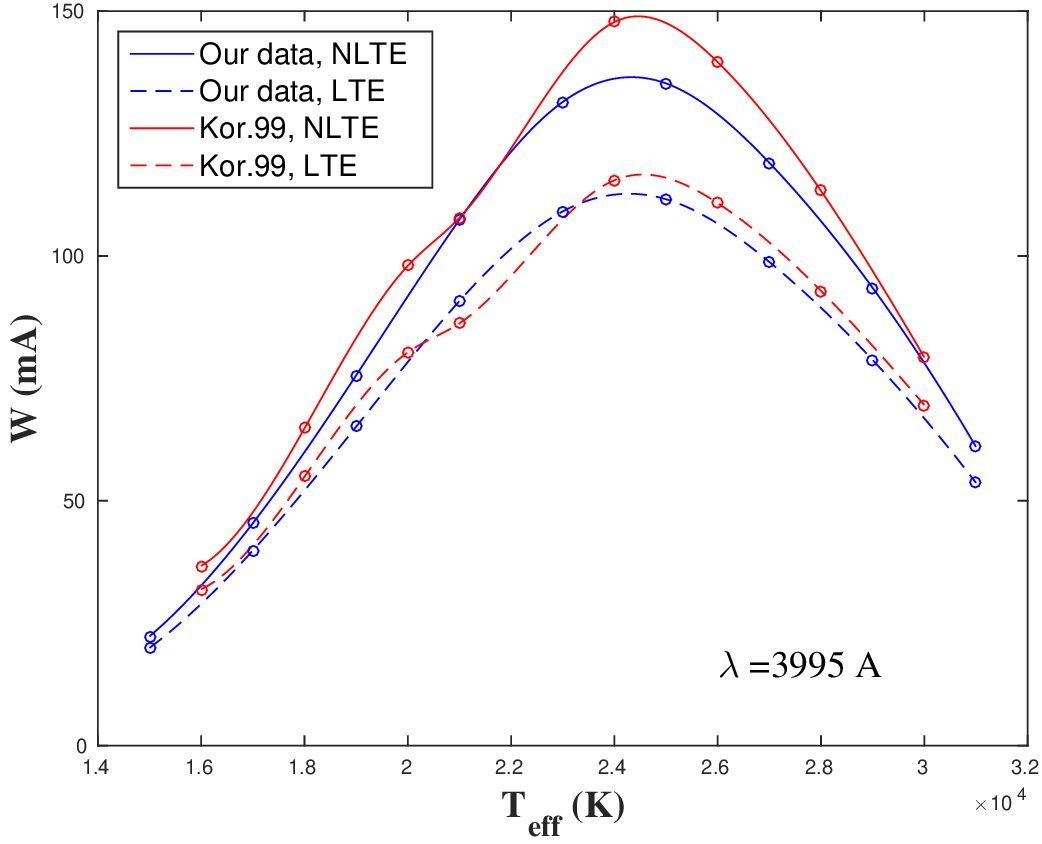}
\label{Equiv_Width_3995}
}     
\hspace{0.3cm}                    
\subfloat[]{
\includegraphics[scale= 0.4]{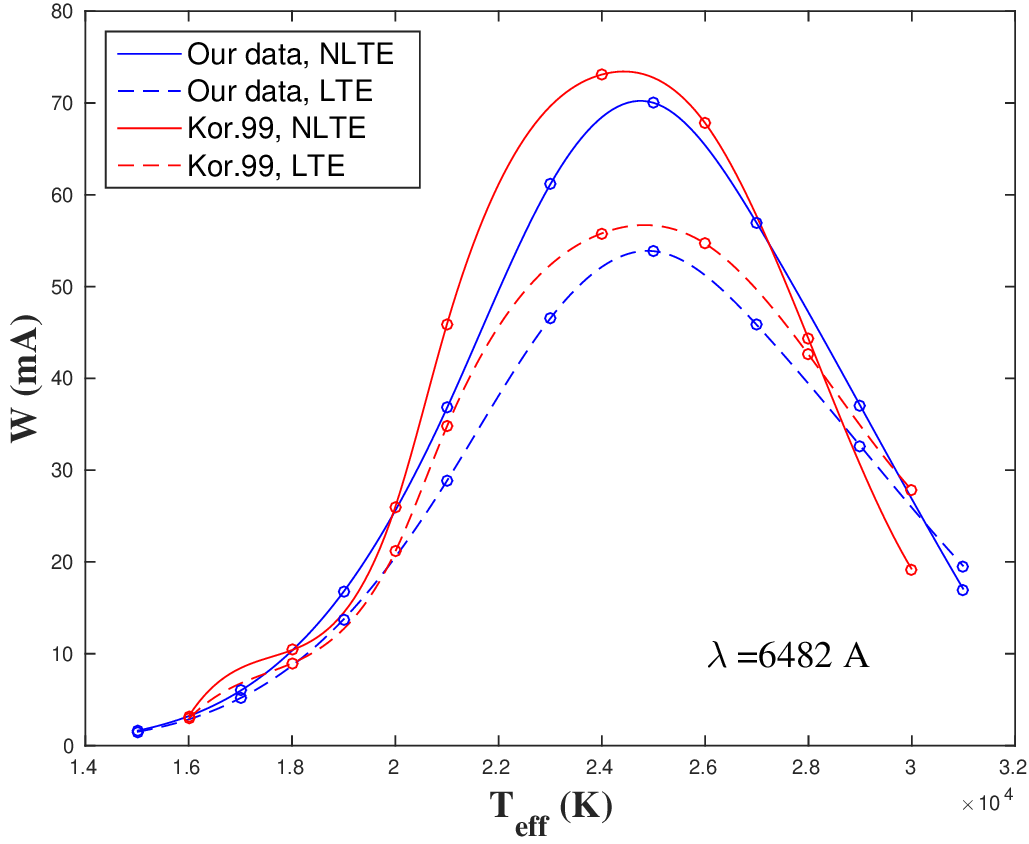}
\label{Equiv_Width_6482}
}
\vspace{0.2cm}
\subfloat[]{
\includegraphics[scale= 0.4]{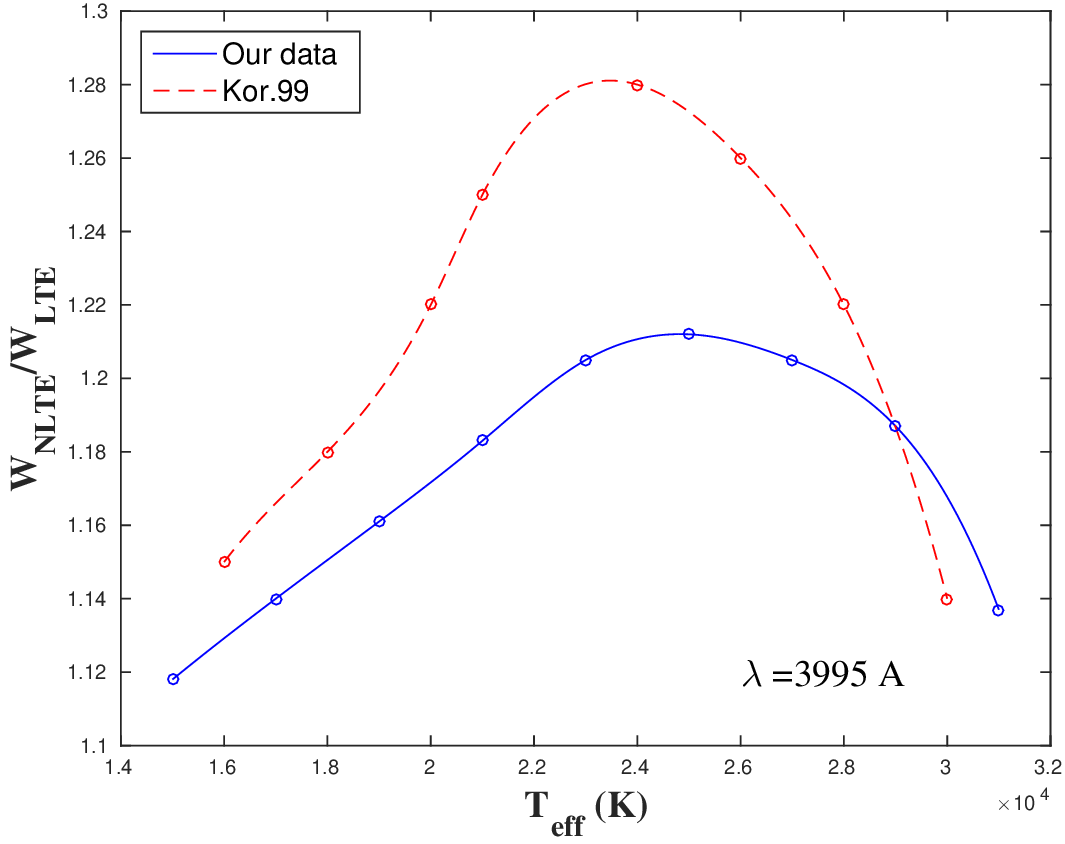}
\label{Equiv_Width_ratio_3995}
}
\hspace{0.3cm}
\vspace{0.2cm}
\subfloat[]{
\includegraphics[scale= 0.4]{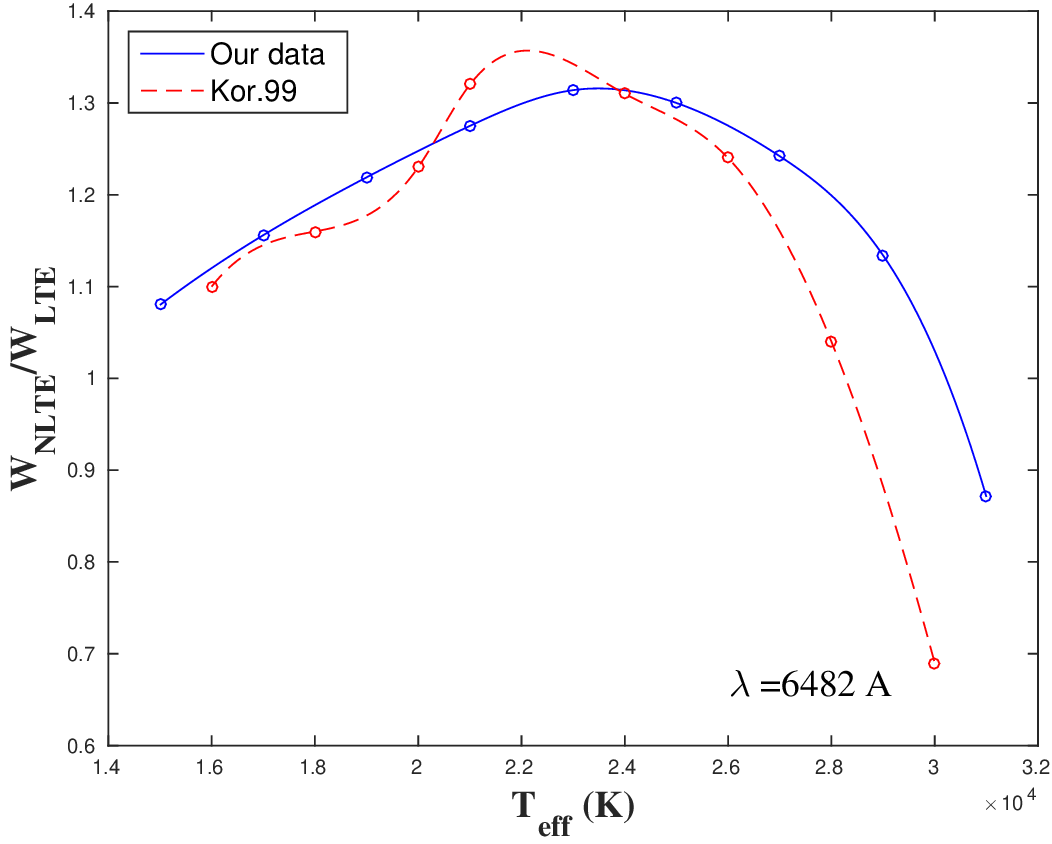}
\label{Equiv_Width_ratio_6482}
}
\caption{The LTE and non-LTE equivalent widths of N\,{\sc ii}
$\lambda\,3995$; (panel~a) and $\lambda\,6482$, (panel~b) respectively,
as a function of $T_{\rm eff}$ for $\log g=$ 4.0, $\xi_t= 5\;\rm km\,s^{-1}$ and $
\epsilon_N=$ 7.95. The blue symbols are the current results and the red
symbols, \citet{kor99}.  The two bottom panels, (c) and (d), give the ratio of
the non-LTE and LTE equivalent widths.  \label{Equiv_Width}}
\end{figure*}

\subsection{ N\,{\sc ii} Equivalent Widths}

Figure~\ref{Equiv_Width} shows the predicted LTE and non-LTE equivalent
widths (in milli-\AA) for N\,{\sc ii} $\lambda\,3995$ and $\lambda\,6482$
representing transitions from levels 9 to 16 $(3s\,^1P^o\rightarrow
3p\,^3D)$ and levels 9 to 11 ($ 3s\,^1P^o \rightarrow 3p\,^1P$),
respectively.  For comparison, the predictions of \citet{kor99}
are also shown. In general, there is reasonable agreement between
the two calculations. For $\lambda\,3995$, we find
smaller deviations from LTE near the line's maximum strength at
$T_{\rm eff}\approx 25,000\,$K (Figure~\ref{Equiv_Width_ratio_3995}).
For $\lambda\,6482$, agreement is good for $T_{\rm eff} < 24,000\,$K;
however, for hotter $T_{\rm eff}$, we again predict smaller departures
from the LTE equivalent width than \citet{kor99} (Figure~\ref{Equiv_Width_ratio_6482}).

\begin{figure*}
\centering
\subfloat[]{
\includegraphics[scale= 0.4]{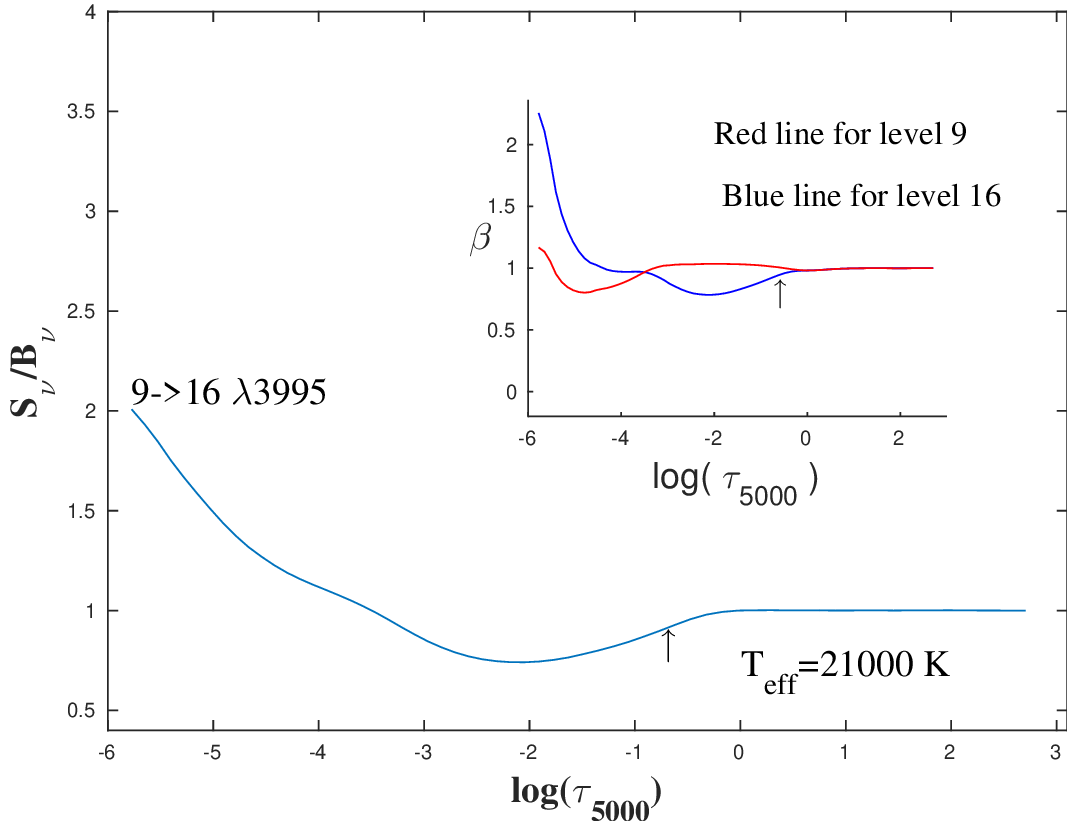}
\label{Snu_3995_T21000}
}
\vspace{0.08cm}
\hspace{0.3cm}
\subfloat[]{
\includegraphics[scale= 0.4]{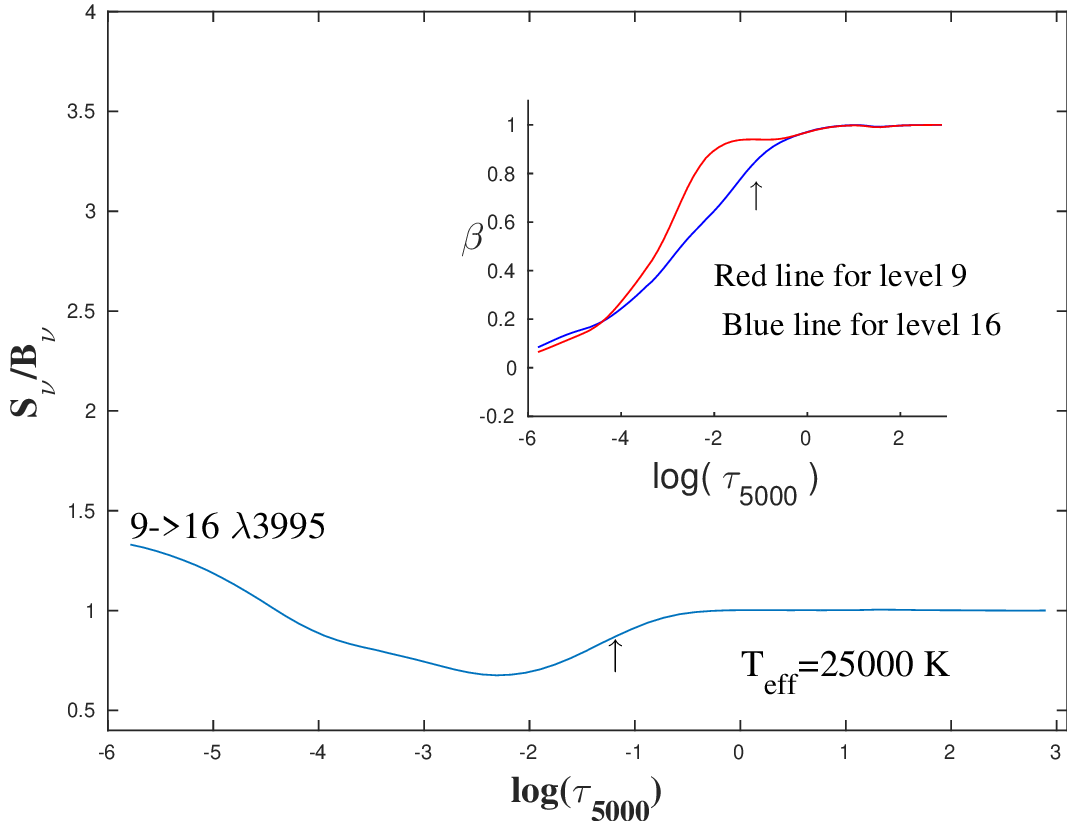}
\label{Snu_3995_T25000}
}
\vspace{0.01cm}
\subfloat[]{
\includegraphics[scale= 0.4]{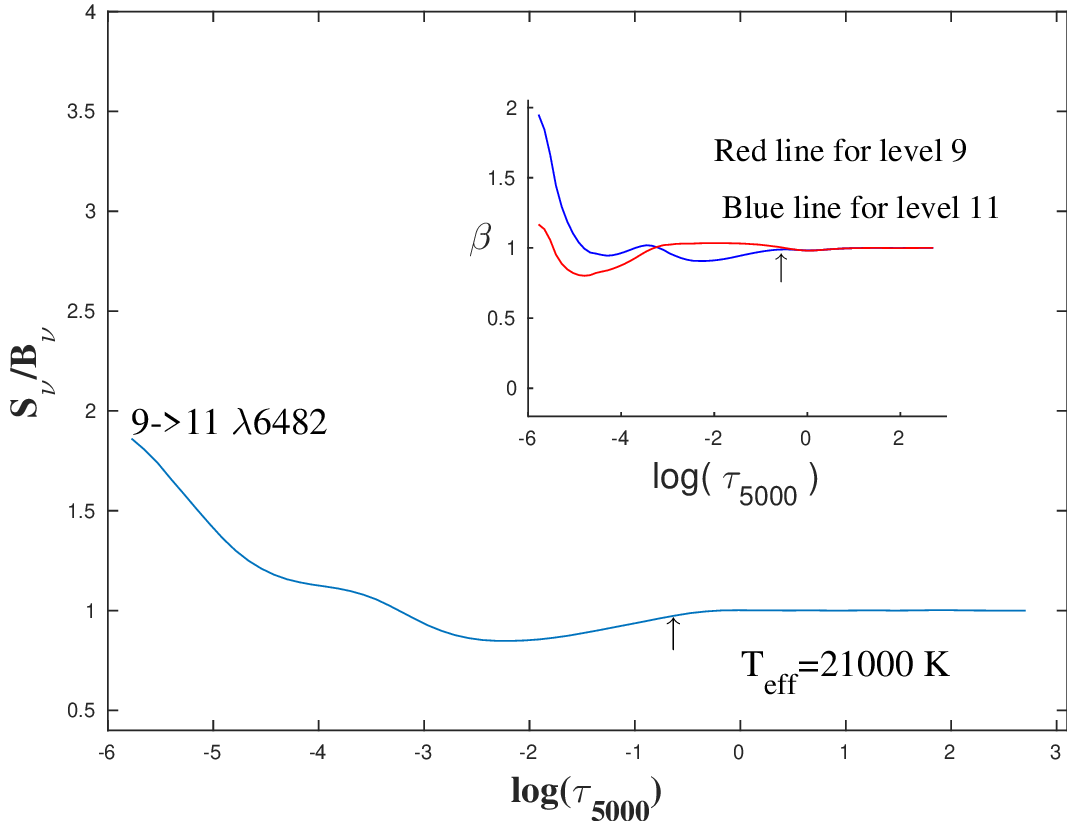}
\label{Snu_6482_T21000}
}
\hspace{0.3cm}
\subfloat[]{
\includegraphics[scale= 0.4]{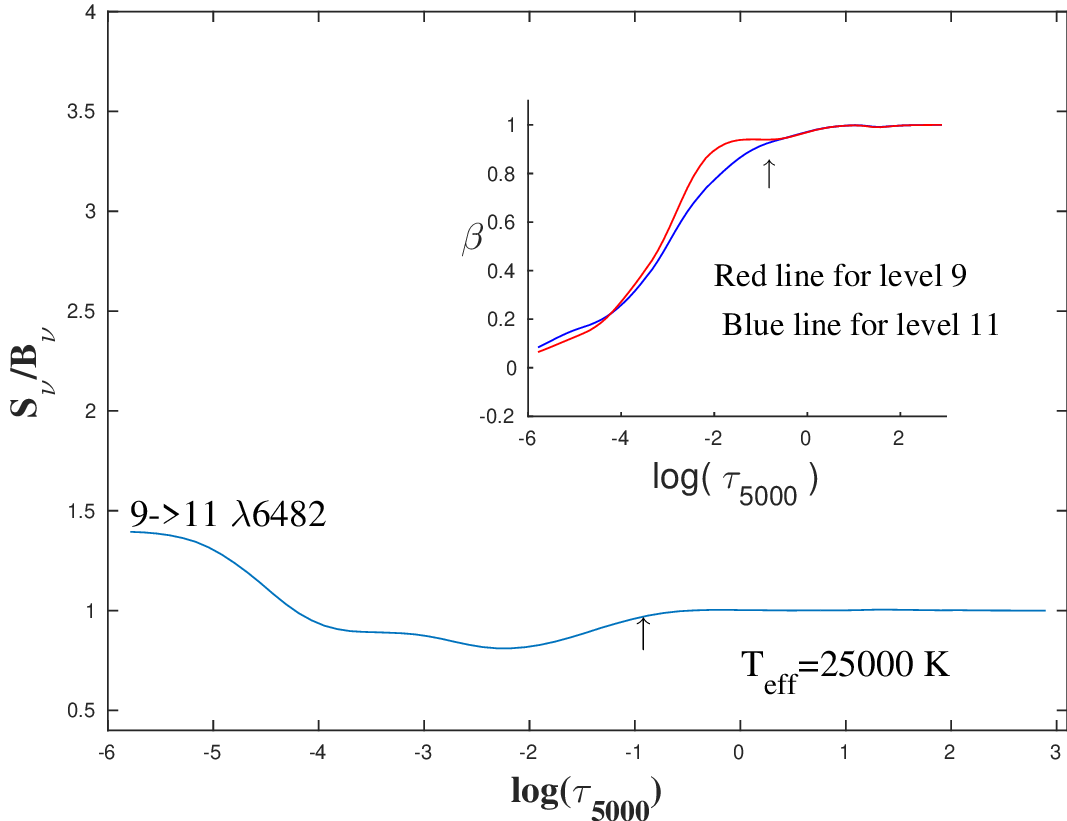}
\label{Snu_6482_T25000}
}
\caption{The line source functions as a function of $\log\tau_{5000}$ 
for N\,{\sc ii} $\lambda\,3995$
and $\lambda\,6482$ at two $T_{\rm eff}$, 21000 and 25000\,K. Both
models assumed $\log g=4.0$, $\xi_t= 5\;\rm km\,s^{-1}$ and the solar nitrogen
abundance.  The departure coefficients of the upper and lower energy
levels are shown in the panel inserts.  The arrows point to the depth
of formation of the line centre flux as defined by the contribution function
of X.}
\label{Source_fun}
\end{figure*}

In general, the non-LTE equivalent widths are predicted to be larger than
the corresponding LTE values. These differences result from deviations of
the line source function from the local Planck function and the non-LTE correction
to the line optical depth scale. In terms of the departure coefficients
of the upper and lower levels, $\beta_u$ and $\beta_l$, the line source function is
\begin{equation}
S_l =\frac{2h\nu^3}{c^2} \left(\frac{\beta_l}{\beta_u}e^{\frac{h\nu}{kT}}-1\right)^{-1}\,,
\end{equation}
and the line optical depth scale is,
\begin{equation}
d\tau^l_{\nu}=-\frac{h\nu}{4\pi}\left(\beta_l\,n^*_l - \beta_u\,n^*_u\right) \, \phi_{\nu}\,dz \,.
\end{equation}
Here $\phi_{\nu}$ is the line absorption profile, $n^*$ are the
LTE level populations, and $dz$ is the physical step (in cm) along
the ray.  Complete redistribution, the equality of the line emission
and absorption profiles, has been assumed \citep{mih78}. Note that if
$h\nu\gg kT$ (i.e.\ the photon energy exceeds the local kinetic energy),
we have the scalings $S_l \propto \beta_u/\beta_l$ and $d\tau^l_{\nu}
\propto \beta_l$. As the Eddington-Barbier relation states that the
emergent intensity is characteristic of the source function at an optical
depth of $\approx 2/3$, we see that $\beta_l$ affects how deeply we see
into the atmosphere, whereas $\beta_u/\beta_l$ controls the value of the
source function at this point.  This emphasizes that even in cases where
$S_l=B_{\nu}(T_{\rm e})$, i.e.\ $\beta_u/\beta_l=1$, there can still be
large non-LTE effects, for example, if $\beta_l=\beta_u \ll 1$.

\begin{figure*}
\centering
\subfloat[]{
\includegraphics[scale= 0.40]{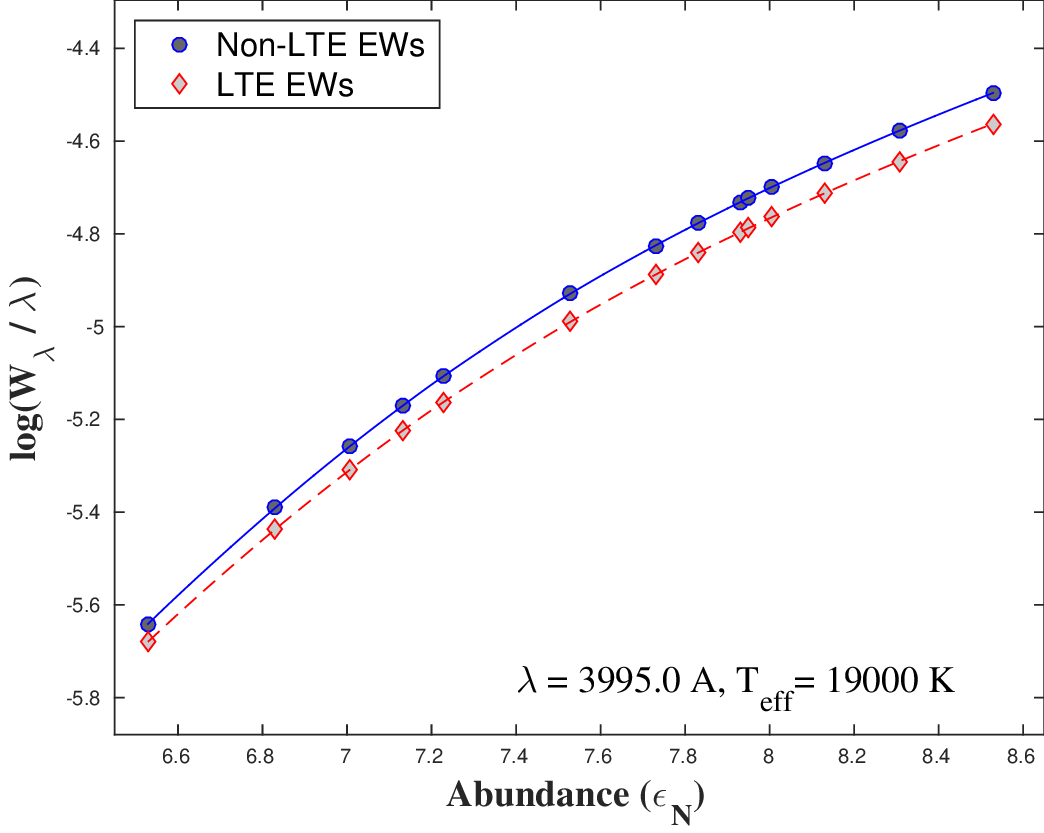}
\label{Growth_Curve__3995A_T21000}
}
\vspace{0.005cm}
\hspace{0.1cm}                    
\subfloat[]{
\includegraphics[scale= 0.40]{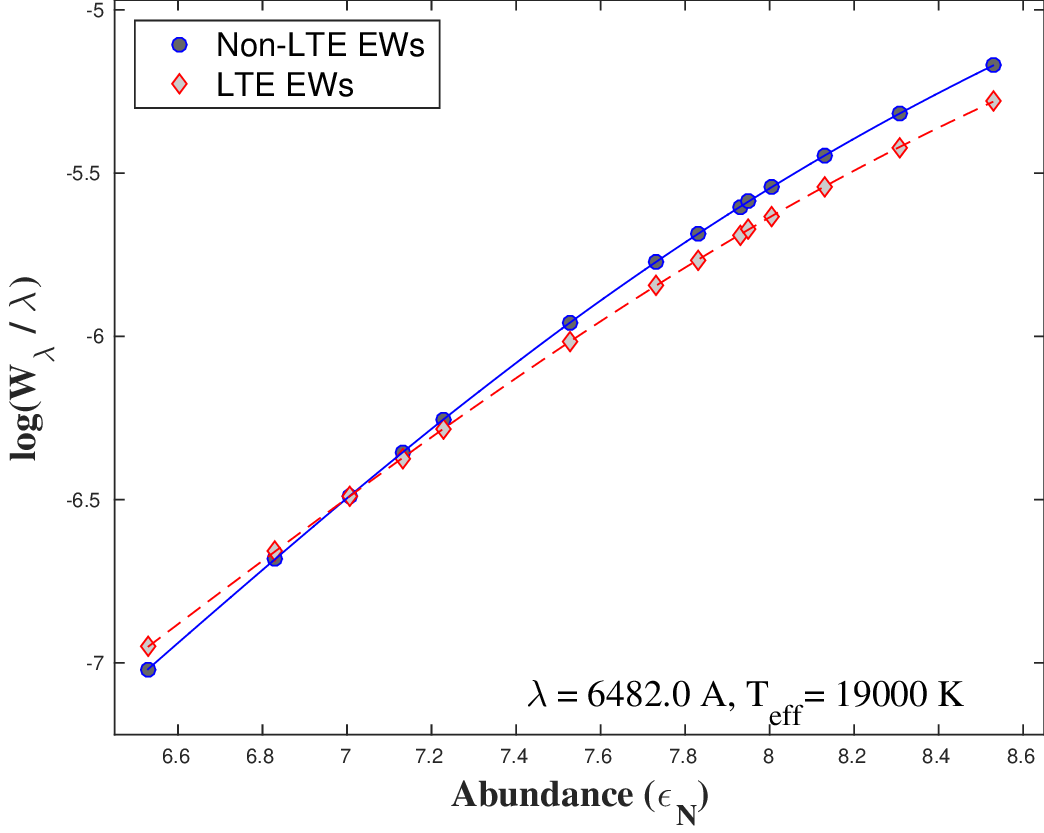}
\label{Growth_Curve__6482A_T21000}
}
\vspace{0.005cm}
\subfloat[]{
\includegraphics[scale= 0.40]{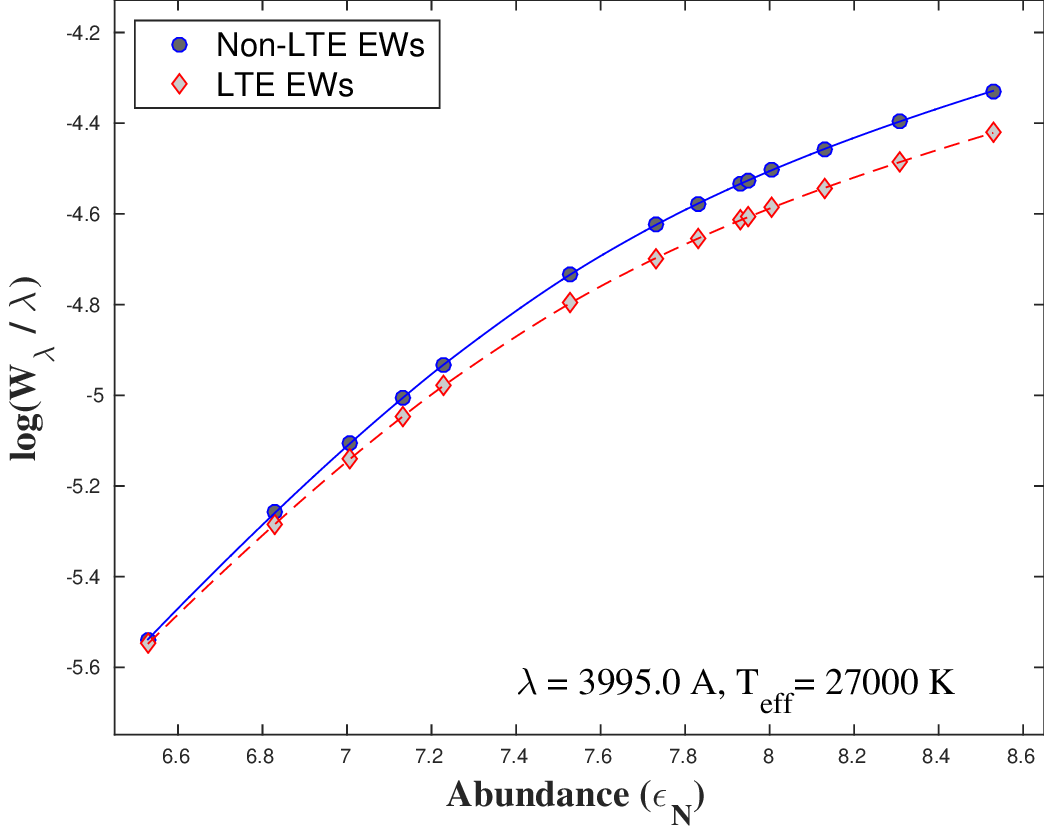}
\label{Growth_Curve__3995A_T25000}
}
\hspace{0.1cm}                    
\subfloat[]{
\includegraphics[scale= 0.40]{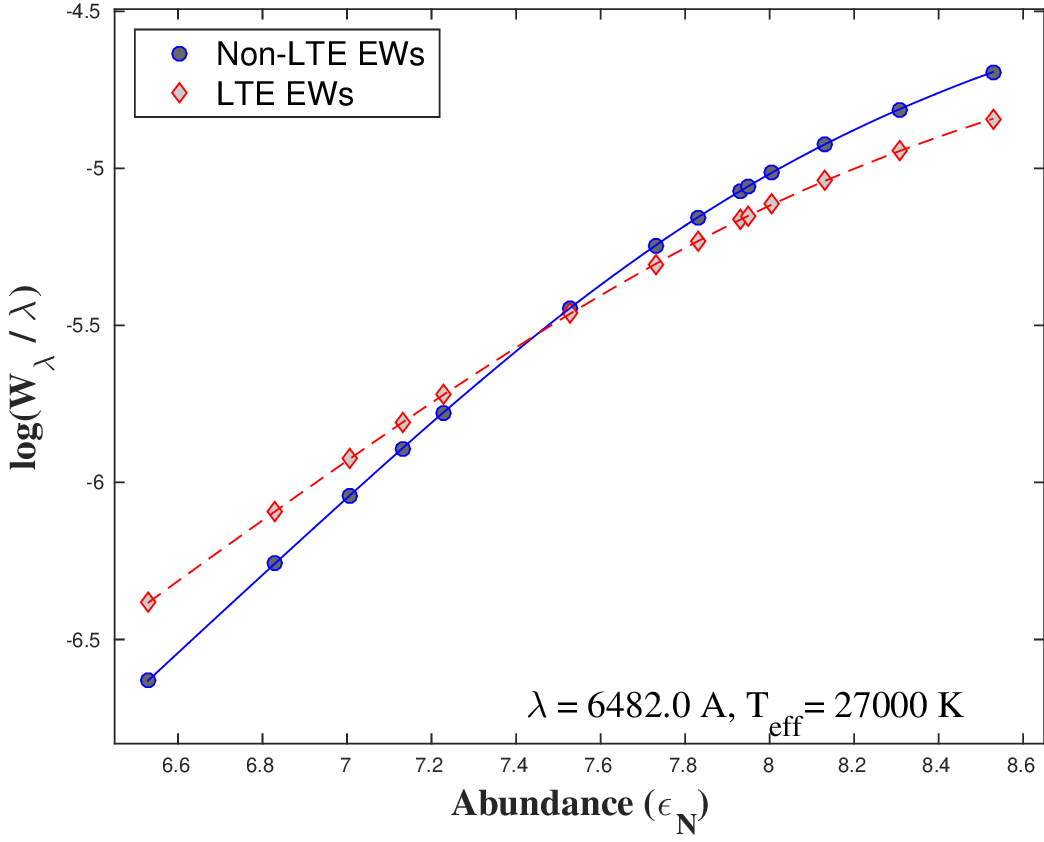}
\label{Growth_Curve__6482A_T25000}
}
\caption{LTE and non-LTE curves of growth for N\,{\sc ii} 
$\lambda\,6482$ and $\lambda\,3995$ as a function of nitrogen abundance,
$\rm \epsilon_N$, at the indicated $T_{\rm eff}$ with
$\log g$= 4.0, and $\xi_t= 5\;\rm km\,s^{-1}$. The blue, solid lines
represent the predicted non-LTE equivalent widths and the dashed, red,
dashed lines represent the corresponding LTE values.}
\label{Growth_Curve}
\end{figure*}

Panels~\subref{Snu_3995_T21000} and \subref{Snu_3995_T25000} of
Figure~\ref{Source_fun} show the ratio of the line source function
to Planck function for N\;{\sc ii} $\lambda\,3995$ as a function of
$\tau_{5000}$ at two $T_{\rm eff}$, 21,000 and 25,000 K.  The departure
coefficients of the upper and lower levels are shown as inserts in
each figure.

Consider first $\lambda\,3995$, transition $9\rightarrow 16$.  The
depth of formation of the line centre flux is marked with an arrow,
following the flux contribution function proposed by \cite{ajn91}.
At $T_{\rm eff}$ of 21,000\;K there is a small overpopulation of the the
ninth energy level and a small under-population of the sixteenth level,
while at $T_{\rm eff}$ of 25,000 K, there is under-population of both
levels. In both cases, the main effect is the larger under-population of
the upper energy level which acts to reduce the line source function in
the line-forming region leading to non-LTE strengthening of the line.
The behaviour of $\lambda\,6482\,$\AA, transition $9\rightarrow 11$, is
qualitatively similar. However, for $T_{\rm eff}$ greater than 28,000\;K,
there is a non-LTE weakening of the line driven by the overionization
of N\,{\sc ii} in such models.  

For both $\lambda\,3995$ and $\lambda\,6482$, non-LTE strengthening
of the lines reaches a maximum near $T_{\rm eff}\approx 25,000\;$K
where the lines are about 20\% and 30\% stronger than the LTE
predictions, respectively (Figures~\ref{Equiv_Width_3995} and
\ref{Equiv_Width_6482}). This overall behaviour is in agreement
with the calculation of \citet{kor99}, although our overall non-LTE
strengthening of $\lambda\,3995$ is less and our non-LTE weakening of
$\lambda\,6482\,$\AA\ for high $T_{\rm eff}$ is also less. The stronger
non-LTE effects seen by \citet{kor99} may reflect their lower number of
bound-bound radiative transitions included, 266 transitions in total,
with 92 included in the linearization procedure and the rest kept at
fixed rates. In the current work, a total of 580 radiative transitions
were included, none with fixed rates, representing all LS transitions
with oscillator strengths greater than or equal to $10^{-3}$.
We confirm that it is the larger number of radiative transitions included in the present work that
explains most of the differences with K99 by constructing a 43~LS level N\,{\sc ii} atom that
includes the same set of allowed rbb transitions as K99. Using this atom,
their non-LTE equivalent widths could be reproduced with a good agreement as a function of
$T_{\rm eff}$ for the model $\log g$= 4.0, $\xi_t= 5\,\rm km\,s^{-1}$ and $\epsilon_N=7.95$, Table (5) in K99.

\begin{figure*}
\centering
\subfloat[]{
\includegraphics[scale= 0.4]{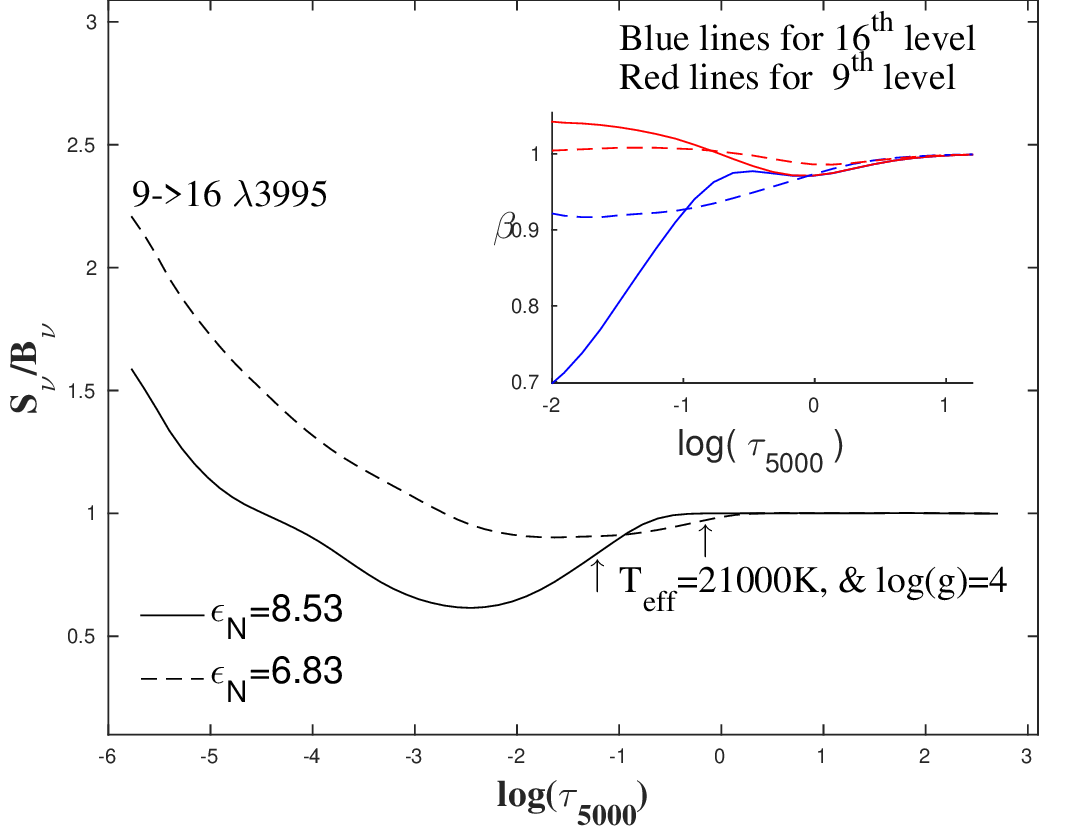}
\label{Snu_3995A_T21000_diff_abund}
}
\hspace{0.3cm}                   
\subfloat[]{
\includegraphics[scale= 0.4]{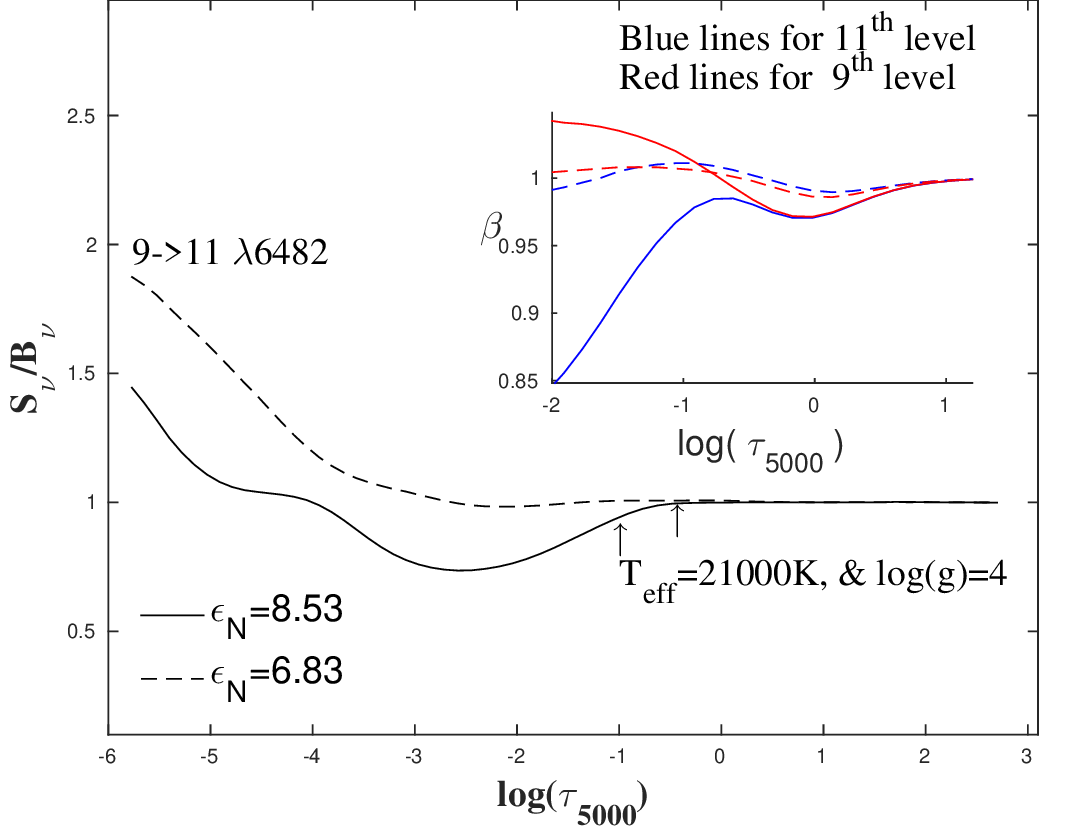}
\label{Snu_6482A_T21000_diff_abund}
}
\vspace{0.1cm}
\subfloat[]{
\includegraphics[scale= 0.4]{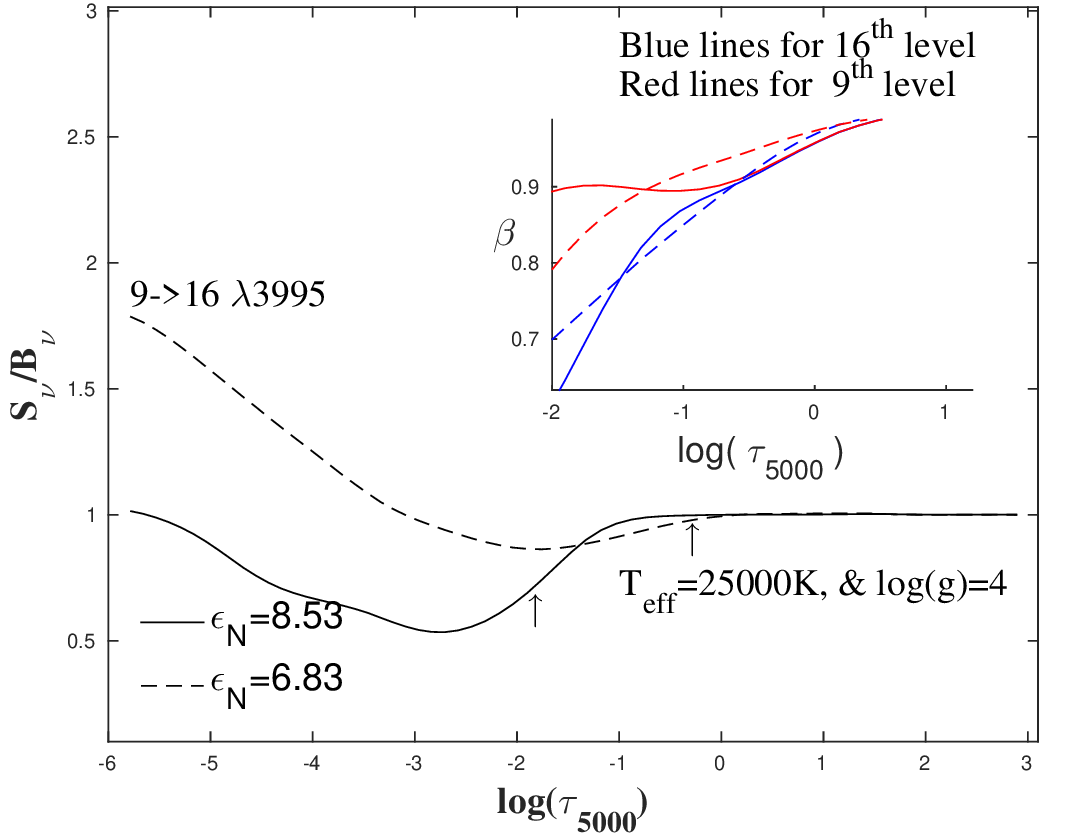}
\label{Snu_3995A_T25000_diff_abund}
}
\vspace{0.1cm}
\hspace{0.3cm}                   
\subfloat[]{
\includegraphics[scale= 0.4]{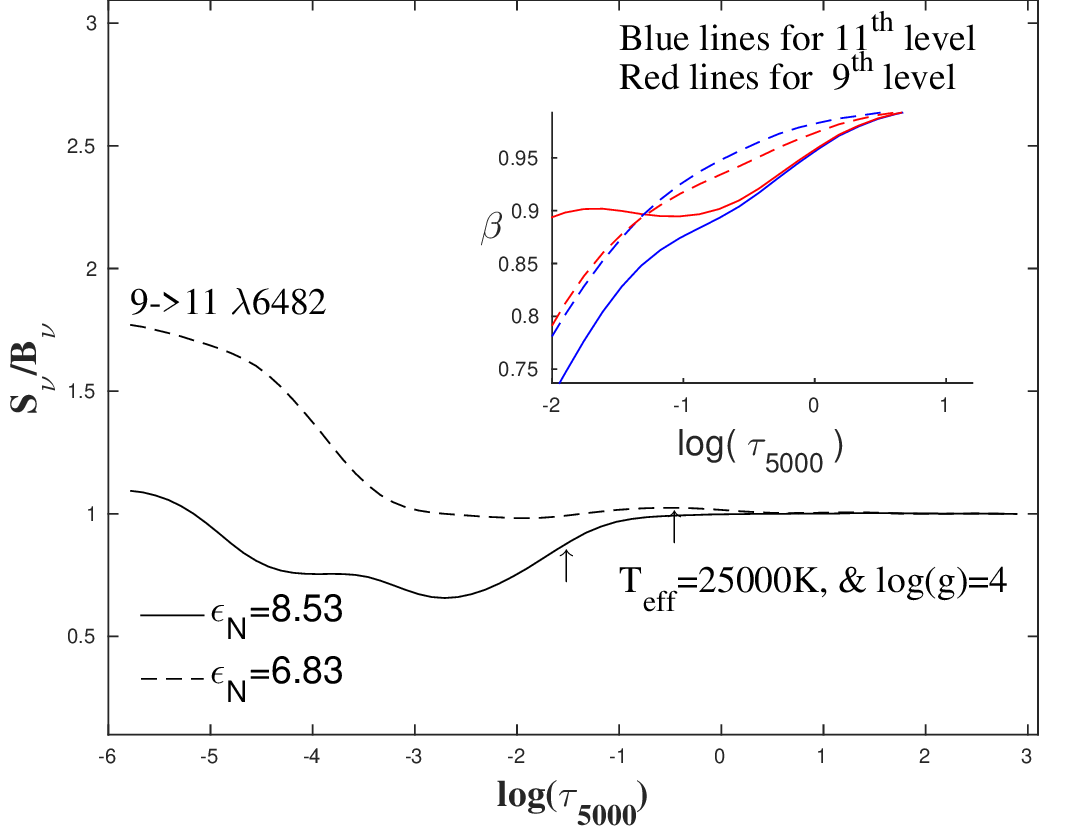}
\label{Snu_6482A_T25000_diff_abund}
}                  
\caption{The source functions of the N\;{\sc ii} $\lambda\,3995$ for two different
nitrogen abundances, $\epsilon_N=6.83$ (solid lines) and $8.53$ (dashed lines). The $T_{\rm eff}$
and $\log\,g$ of the models are as indicated. The microturbulent velocity was
$\xi_t=5\;\rm km\,s^{-1}$.}
\label{Snu_diff_abund}
\end{figure*}

There is also a dependence of the size of the predicted non-LTE effects on
the nitrogen abundance. This is illustrated in Figure~\ref{Growth_Curve}
which show curves of growth, $\log (W_\lambda/\lambda)$ versus
$\epsilon_N$, for both $\lambda\,3995$ and $\lambda\,6482$ at two
$T_{\rm eff}$, 19,000\;K and 27,000\;K. A gravity of $\log\,g=4.0$ and
a microturbulent velocity of $\xi_t=5\;\rm km\,s^{-1}$ were assumed for
all examples.  For $\lambda\,3995$, there is non-LTE strengthening at
all nitrogen abundances, with the largest effect at the highest abundance
considered. For $\lambda\,6482$, there is a reversing trend with non-LTE
weakening predicted for small abundances and a non-LTE strengthening
at larger abundances. The transition occurs at abundances somewhat less
than the solar value.

This abundance effect is further explored in Figure~\ref{Snu_diff_abund}
which shows the line source functions and departure coefficients
of both $\lambda\,3995$ and $\lambda\,6482$ at the two extreme
values of the nitrogen abundance considered, $\epsilon_N=6.83$ ($-1.0\;$dex
relative to solar) and $\epsilon_N=8.53$ ($+0.7\;$dex relative to solar).
The figure shows that the increase of the nitrogen abundance reduces the line source function value and shifts the line formation regions to smaller optical depths ($\log \tau_{5000}$).

Grids of non-LTE equivalent widths for $\lambda\,3995$
and $\lambda\,6482$ over all $T_{\rm eff}$, $\log\,g$ and
$\epsilon_N$ considered are given in Tables~\ref{multi_wv3995_vturb5_2}
and \ref{multi_wv6482_zeta5}. The microturbulent velocity was set to
$5\;\rm km\,s^{-1}$. Full grids of equivalent widths for all transitions
of Table~\ref{rbb-trans-data} over all models and microturbulences
considered are available on-line.

Finally we note that the equivalent width of $\lambda\,6482$ at
the highest temperature considered in Table~\ref{multi_wv6482_zeta5},
31,000\;K, becomes weakly negative, indicating line emission. This is a
well-known non-LTE effect that can occur when $h\nu/kT\ll 1$: the source
function becomes very sensitive to small departures from LTE and can
rise in the outer layers, even though the photospheric temperature falls
with height.  This effect is extensively discussed in \cite{car92_2}
and \cite{sig96b}.

\subsection{A Multi-MULTI Analysis}
\label{MM-analysis}

\begin{figure*}
\centering
\subfloat[]{
\includegraphics[scale= 0.4]{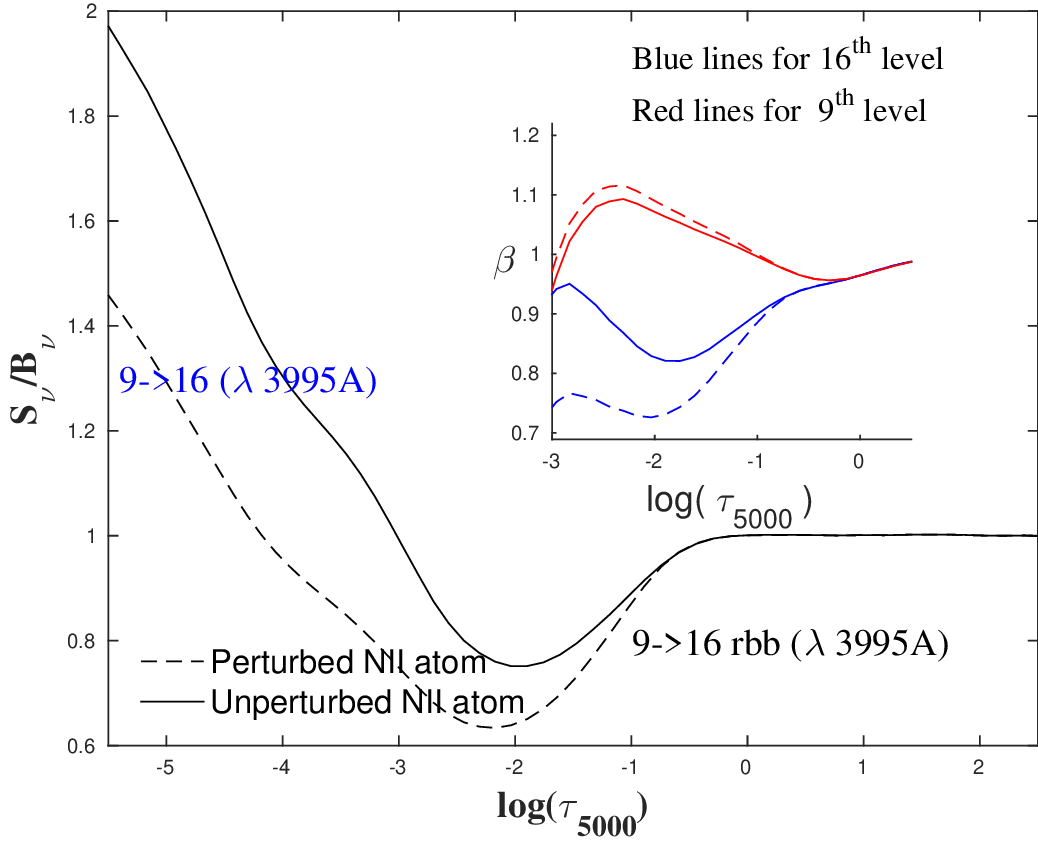}
\label{Snu_3995_pret1}
}
\vspace{0.005cm}
\hspace{0.3cm}
\subfloat[]{
\includegraphics[scale= 0.4]{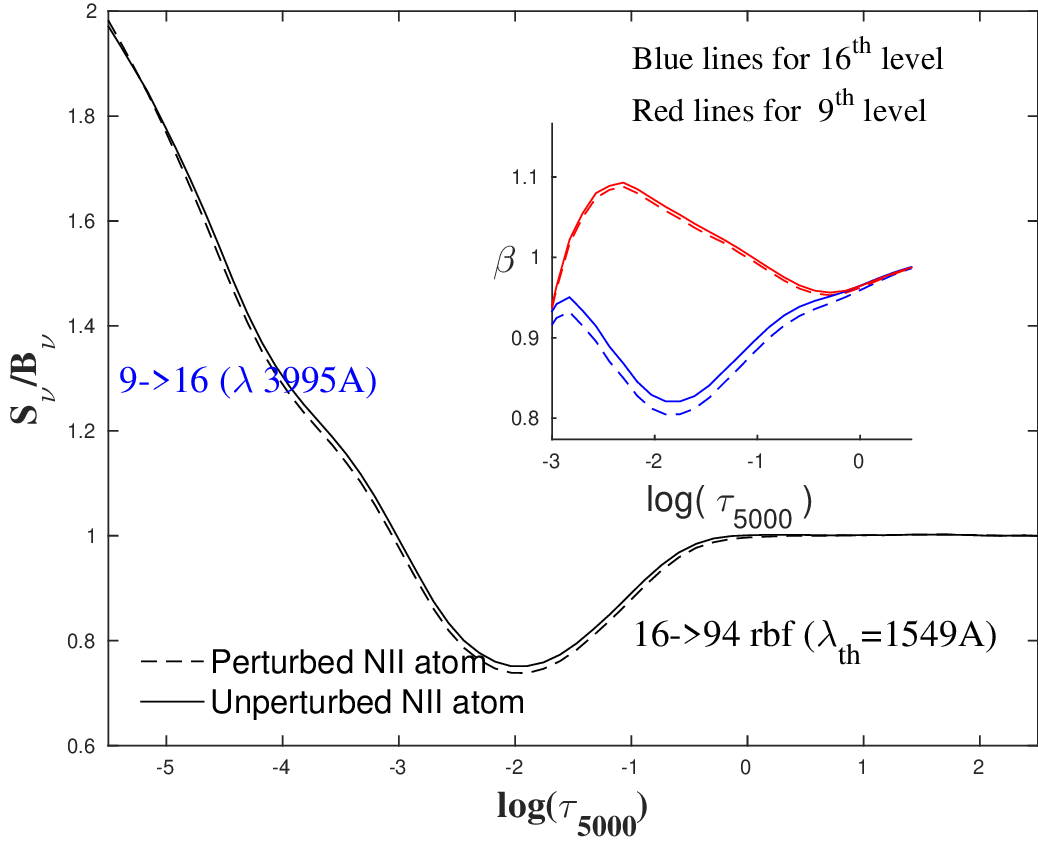}
\label{Snu_3995_pret2}
}
\vspace{0.005cm}
\subfloat[]{
\includegraphics[scale= 0.4]{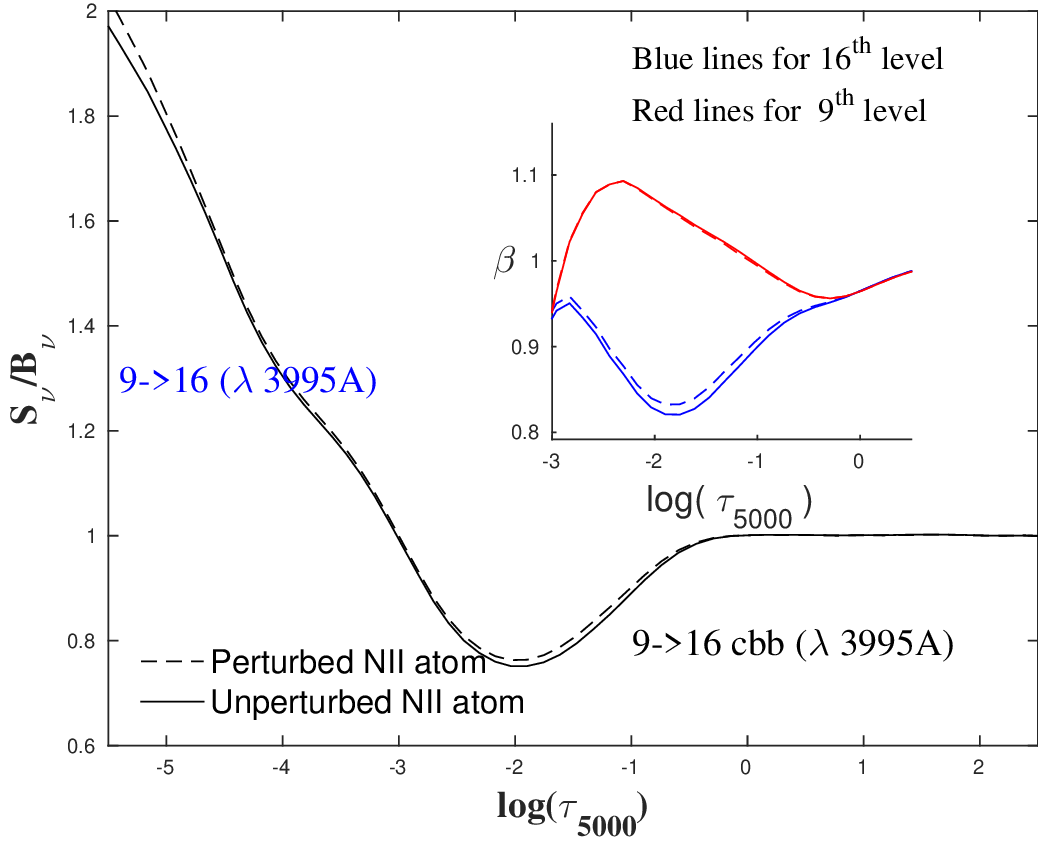}
\label{Snu_3995_pret3}
}
\hspace{0.3cm}
\subfloat[]{
\includegraphics[scale= 0.4]{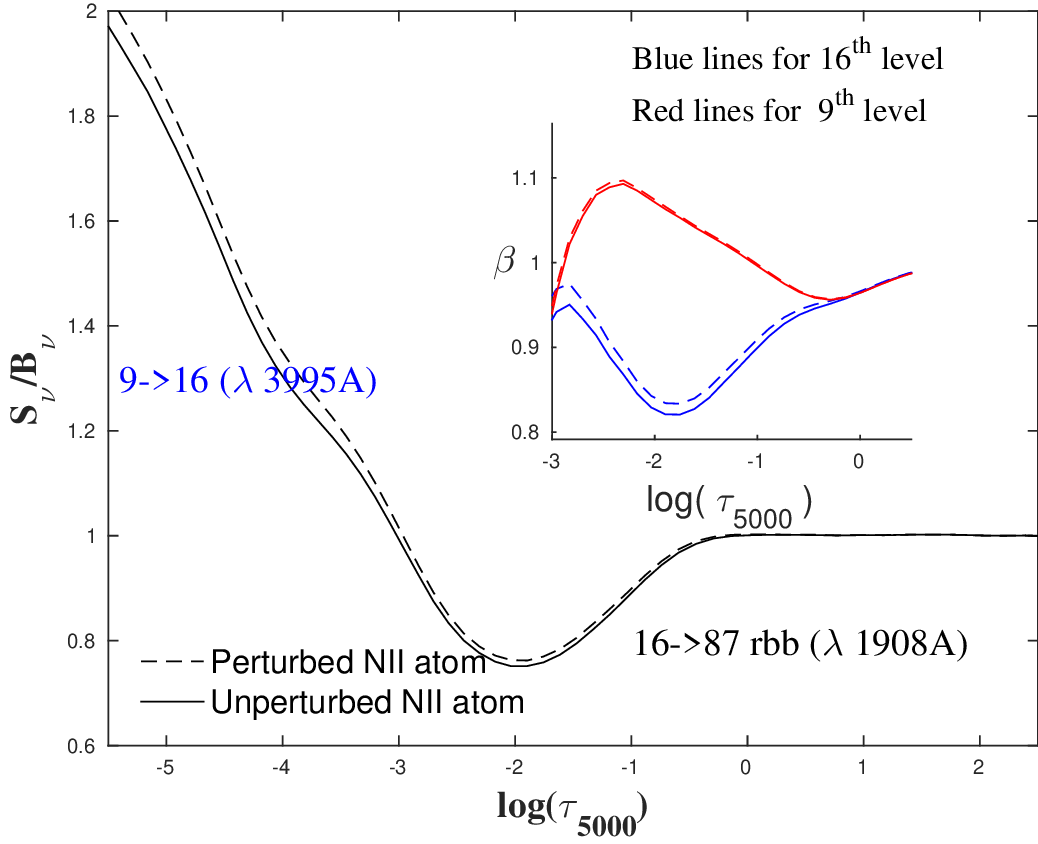}
\label{Snu_3995_pret4}
}
\caption{The line source function of N\;{\sc ii} $\lambda\,3995$ line for the perturbed nitrogen atom and the departure coefficients of the upper and lower energy levels of this transition (dashed lines) for $\rm T_{eff}=$ 23000 K,  $\log g=$ 4.0, $\xi_t= 5\;\rm km\,s^{-1}$, and $\rm \epsilon_{N, solar}$= 7.83. For comparison the results of the unperturbed atom were added (solid lines), where the perturbed transition is displayed in the lower right corner of each panel.}
\label{Source_func_preturbed_atom}
\end{figure*}

In order to investigate which of the radiative and collisional
transitions included in the atom have the most significant effects on
the predicted equivalent widths of the N\,{\sc ii} lines of interest,
a series of multi-MULTI analysis were carried out \citep{car92_2}. In
a multi-MULTI analysis, a single radiative or collisional transition
is perturbed by doubling its rate, and a new converged solution is
obtained for the perturbed atom. The predicted equivalent widths from
this new converged solution are compared with the reference solution of
the unperturbed atom. Table~\ref{multi-multi_23kK} shows the top ten
radiative/collisional transitions affecting the predicted equivalent
widths of $\lambda\,6482$ and $\lambda\,3995$ for $\rm
T_{eff}\,$= 23,000 K, $\log g$= 4.0, $\xi_t\,=$5 $\rm km\,s^{-1}$,  and a solar
nitrogen abundance.  Corresponding to Table~\ref{multi-multi_23kK},
Figure~\ref{Snu_3995_pret1} shows the effect on the line source function
and upper and lower level departure coefficients of the top four rates
for $\lambda\,3995$.

\begin{table}
\caption{multi-MULTI Analysis at $\rm T_{eff}$= 23000 K, $\log g$=4.0, and $\xi_t= 5.0\;\rm km\,s^{-1}$
\label{multi-multi_23kK}}
\begin{center}
{\small
\begin{tabular*}{0.49\textwidth}{@{}l @{}l @{}l @{}l @{}l @{}l @{}l l@{}}
\hline\hline
Transition & \multicolumn{5}{c}{Perturbed transition} & Type & $\%$\\
\hline
3995.0\,\AA &   9& $\rightarrow $& 16& \multicolumn{1}{l}{($ 2p \;3s \;^1P^o$} &  \multicolumn{1}{l}{-$ 2p \;3p \;^1D$)} & \multicolumn{1}{l}{rbb} &35.22\\
&   16& $\rightarrow $& 94& \multicolumn{1}{l}{($ 2p \;3p \;^1D$} &  \multicolumn{1}{l}{-$ 2s^2\;2p\;^2P^o$ N\,{\sc iii})} & \multicolumn{1}{l}{rbf} &     1.67\\
&   9& $\rightarrow $& 16& \multicolumn{1}{l}{($ 2p \;3s \;^1P^o$} &  \multicolumn{1}{l}{-$ 2p \;3p \;^1D$)} & \multicolumn{1}{l}{cbb} &    -1.43\\
&   16& $\rightarrow $& 87& \multicolumn{1}{l}{($ 2p \;3p \;^1D$} &  \multicolumn{1}{l}{-$\rm high\;l,n=6 $)} & \multicolumn{1}{l}{rbb} &    -1.29\\
&   16& $\rightarrow $& 22& \multicolumn{1}{l}{($ 2p \;3p \;^1D$} &  \multicolumn{1}{l}{-$ 2p \;3d \;^1F^o$)}  & \multicolumn{1}{l}{rbb} &    -1.21\\
&  22& $\rightarrow $& 87& \multicolumn{1}{l}{($ 2p \;3d \;^1F^o$} &  \multicolumn{1}{l}{-$\rm high\;l,n=6 $)} & \multicolumn{1}{l}{rbb} &    -0.94\\
&   8& $\rightarrow $& 94& \multicolumn{1}{l}{($ 2p \;3s \;^3P^o$} &  \multicolumn{1}{l}{-$ 2s^2\;2p\;^2P^o$ N\,{\sc iii})} & \multicolumn{1}{l}{rbf} &    -0.79\\
&   16& $\rightarrow $& 25& \multicolumn{1}{l}{($ 2p \;3p \;^1D$} &  \multicolumn{1}{l}{-$ 2p \;4s \;^1P^o$)}  & \multicolumn{1}{l}{rbb} &    -0.75\\
&    1& $\rightarrow $& 94& \multicolumn{1}{l}{($ 2p^2\;^3P$} &  \multicolumn{1}{l}{-$ 2s^2\;2p\;^2P^o$ N\,{\sc iii})} & \multicolumn{1}{l}{rbf} &     0.71\\
&   12& $\rightarrow $& 16& \multicolumn{1}{l}{($ 2p \;3p \;^3D$} &  \multicolumn{1}{l}{-$ 2p \;3p \;^1D$)} & \multicolumn{1}{l}{cbb} &    -0.68\\ \cline{2-8}
6482.0\,\AA &   9& $\rightarrow $& 11& \multicolumn{1}{l}{($ 2p \;3s \;^1P^o$} &  \multicolumn{1}{l}{-$ 2p \;3p \;^1P$)} & \multicolumn{1}{l}{rbb} &    71.67\\
&   11& $\rightarrow $& 19& \multicolumn{1}{l}{($ 2p \;3p \;^1P$} &  \multicolumn{1}{l}{-$ 2p \;3d \;^1D^o$)}  & \multicolumn{1}{l}{rbb} &    -6.45\\
&   11& $\rightarrow $& 94& \multicolumn{1}{l}{($ 2p \;3p \;^1P$} &  \multicolumn{1}{l}{-$ 2s^2\;2p\;^2P^o$ N\,{\sc iii})} & \multicolumn{1}{l}{rbf} &     3.95\\
&   9& $\rightarrow $& 16& \multicolumn{1}{l}{($ 2p \;3s \;^1P^o$} &  \multicolumn{1}{l}{-$ 2p \;3p \;^1D$)} & \multicolumn{1}{l}{rbb} &     3.41\\
&   9& $\rightarrow $& 11& \multicolumn{1}{l}{($ 2p \;3s \;^1P^o$} &  \multicolumn{1}{l}{-$ 2p \;3p \;^1P$)} & \multicolumn{1}{l}{cbb} &    -3.12\\
&   11& $\rightarrow $& 23& \multicolumn{1}{l}{($ 2p \;3p \;^1P$} &  \multicolumn{1}{l}{-$ 2p \;3d \;^1P^o$)}  & \multicolumn{1}{l}{rbb} &    -2.95\\
&  19& $\rightarrow $& 87& \multicolumn{1}{l}{($ 2p \;3d \;^1D^o$} &  \multicolumn{1}{l}{-$\rm high\;l,n=6 $)} & \multicolumn{1}{l}{rbb} &    -2.91\\
&   7& $\rightarrow $& 11& \multicolumn{1}{l}{($2p^3 \;^1D^o$} &  \multicolumn{1}{l}{-$ 2p \;3p \;^1P$)} & \multicolumn{1}{l}{rbb} &     2.92\\
&    1& $\rightarrow $&  6& \multicolumn{1}{l}{($ 2p^2 \;^3P$} &  \multicolumn{1}{l}{-$2p^3 \;^3P^o$)}  & \multicolumn{1}{l}{rbb} &    -2.03\\
&    1& $\rightarrow $&  6& \multicolumn{1}{l}{($ 2p^2 \;^3P$} &  \multicolumn{1}{l}{-$2p^3 \;^3P^o$)}  & \multicolumn{1}{l}{cbb} &     2.00\\
\hline
\end{tabular*}}
\end{center}
\noindent{ {\small Note: rbb and rbf refer to bound-bound and bound-free radiative transitions, respectively, and cbb and cbf refer to bound-bound and bound-free collisional transitions, respectively.}}
\end{table}

For $\lambda\,3995$, the equivalent width is most sensitive to its
own radiative transition rate controlled by the oscillator strength;
e.g. doubling  the oscillator strength of the $\lambda\,3995$ line
leads to $\approx$ 40\% increase in the predicted equivalent width. The
increased oscillator strengths shifts the depth of formation of the
line to smaller optical depths ($\log\tau_{5000}$) where the line
source function is lower, leading to an increase in the line strength.
Next in importance was the photoionization rate from the upper level
of $\lambda\,3995$ (level 16). An increase in this rate by a factor of
two leads to an increase in the predicted non-LTE equivalent width of
$\approx\,2$\%.  Increased photoionization from the upper level again
acts to reduce the line source function in the line-forming regions.
The collisional bound-bound transition between the lower and upper
energy levels of the $\lambda\,3995$ (the ninth energy level, and the
sixteenth energy level) is next in importance. Doubling the strength
of this collisional bound-bound transition increases the collisional
coupling between these two levels, and increased collisional coupling
tends to force LTE, i.e. $S_\nu$ comes closer to $B_\nu$ in Figure
\ref{Snu_3995_pret3}; this raises the source function in the line forming
region and therefore the line is weaker.

\begin{table}
\begin{center}
\centering{
\caption{multi-MULTI Analysis at $\rm T_{eff}$= 15000 K, $ \log g$=4.0, and $\xi_t= 5.0\;\rm km\,s^{-1}$
\label{multi-multi_15kK}}
{\small
\begin{tabular*}{0.45\textwidth}{@{}l @{}l @{}l @{}l @{}l @{}l @{}l l@{}}
\hline\hline
Transition & \multicolumn{5}{c}{Perturbed transition} & Type & $\%$\\
\hline
3995.0\,\AA &   9& $\rightarrow $& 16& \multicolumn{1}{l}{($ 2p \;3s \;^1P^o$} &  \multicolumn{1}{l}{-$ 2p \;3p \;^1D$)} & \multicolumn{1}{l}{rbb} &    59.47\\
&   9& $\rightarrow $& 16& \multicolumn{1}{l}{($ 2p \;3s \;^1P^o$} &  \multicolumn{1}{l}{-$ 2p \;3p \;^1D$)} & \multicolumn{1}{l}{cbb} &    -1.35\\
&   12& $\rightarrow $& 16& \multicolumn{1}{l}{($ 2p \;3p \;^3D$} &  \multicolumn{1}{l}{-$ 2p \;3p \;^1D$)} & \multicolumn{1}{l}{cbb} &    -0.75\\
&   8& $\rightarrow $& 16& \multicolumn{1}{l}{($ 2p \;3s \;^3P^o$} &  \multicolumn{1}{l}{-$ 2p \;3p \;^1D$)} & \multicolumn{1}{l}{rbb} &     0.52\\
&   11& $\rightarrow $& 16& \multicolumn{1}{l}{($ 2p \;3p \;^1P$} &  \multicolumn{1}{l}{-$ 2p \;3p \;^1D$)} & \multicolumn{1}{l}{cbb} &    -0.50\\
&   8& $\rightarrow $& 16& \multicolumn{1}{l}{($ 2p \;3s \;^3P^o$} &  \multicolumn{1}{l}{-$ 2p \;3p \;^1D$)} & \multicolumn{1}{l}{cbb} &    -0.43\\
&   16& $\rightarrow $& 22& \multicolumn{1}{l}{($ 2p \;3p \;^1D$} &  \multicolumn{1}{l}{-$ 2p \;3d \;^1F^o$)}  & \multicolumn{1}{l}{rbb} &    -0.35\\
&   16& $\rightarrow $& 25& \multicolumn{1}{l}{($ 2p \;3p \;^1D$} &  \multicolumn{1}{l}{-$ 2p \;4s \;^1P^o$)}  & \multicolumn{1}{l}{rbb} &    -0.33\\
&   16& $\rightarrow $& 22& \multicolumn{1}{l}{($ 2p \;3p \;^1D$} &  \multicolumn{1}{l}{-$ 2p \;3d \;^1F^o$)}  & \multicolumn{1}{l}{cbb} &    -0.33\\
&   15& $\rightarrow $& 16& \multicolumn{1}{l}{($ 2p \;3p \;^3P$} &  \multicolumn{1}{l}{-$ 2p \;3p \;^1D$)} & \multicolumn{1}{l}{cbb} &    -0.25\\ \cline{2-8}
6482.0\,\AA &   9& $\rightarrow $& 11& \multicolumn{1}{l}{($ 2p \;3s \;^1P^o$} &  \multicolumn{1}{l}{-$ 2p \;3p \;^1P$)} & \multicolumn{1}{l}{rbb} &   105.76\\
&   11& $\rightarrow $& 14& \multicolumn{1}{l}{($ 2p \;3p \;^1P$} &  \multicolumn{1}{l}{-$ 2p \;3p \;^3S$)} & \multicolumn{1}{l}{cbb} &    -4.30\\
&   11& $\rightarrow $& 19& \multicolumn{1}{l}{($ 2p \;3p \;^1P$} &  \multicolumn{1}{l}{-$ 2p \;3d \;^1D^o$)}  & \multicolumn{1}{l}{rbb} &    -3.87\\
&   8& $\rightarrow $& 14& \multicolumn{1}{l}{($ 2p \;3s \;^3P^o$} &  \multicolumn{1}{l}{-$ 2p \;3p \;^3S$)} & \multicolumn{1}{l}{rbb} &     3.61\\
&   9& $\rightarrow $& 16& \multicolumn{1}{l}{($ 2p \;3s \;^1P^o$} &  \multicolumn{1}{l}{-$ 2p \;3p \;^1D$)} & \multicolumn{1}{l}{rbb} &     3.27\\
&    1& $\rightarrow $&  6& \multicolumn{1}{l}{($ 2p^2 \;^3P$} &  \multicolumn{1}{l}{-$2p^3 \;^3P^o$)}  & \multicolumn{1}{l}{rbb} &    -2.15\\
&    1& $\rightarrow $&  6& \multicolumn{1}{l}{($ 2p^2 \;^3P$} &  \multicolumn{1}{l}{-$2p^3 \;^3P^o$)}  & \multicolumn{1}{l}{cbb} &     2.15\\
&   11& $\rightarrow $& 16& \multicolumn{1}{l}{($ 2p \;3p \;^1P$} &  \multicolumn{1}{l}{-$ 2p \;3p \;^1D$)} & \multicolumn{1}{l}{cbb} &     2.06\\
&   8& $\rightarrow $& 12& \multicolumn{1}{l}{($ 2p \;3s \;^3P^o$} &  \multicolumn{1}{l}{-$ 2p \;3p \;^3D$)} & \multicolumn{1}{l}{rbb} &     1.89\\
&   7& $\rightarrow $& 11& \multicolumn{1}{l}{($2p^3 \;^1D^o$} &  \multicolumn{1}{l}{-$ 2p \;3p \;^1P$)} & \multicolumn{1}{l}{rbb} &     1.72\\
\hline
\end{tabular*}}}
\end{center}
\end{table}

Similar results were obtained for the multi-MULTI analysis of the
$\lambda\,6482$. Doubling the oscillator strength of the radiative
transition itself causes an increase in the equivalent width by $\approx$
60\%, while doubling the collision strength of this transition reduces
the predicted non-LTE line strengthening and the non-LTE equivalent width
by  $\approx$ 2\%.  

Tables~\ref{multi-multi_15kK} and \ref{multi-multi_29kK} show the results
of a multi-MULTI analysis for the same two N\,{\sc ii} transitions but at
$T_{\rm eff}= 15,000$ and $29,000\;$K.  At $\rm T_{eff}= 15,000\;$K, the
radiative bound-free transitions play little important role as N\,{\sc ii}
is the dominant ionization stage in the line forming region and changes
in the radiative bound-free (rbf) rates have only have only a minor
effects on the N\,{\sc ii} populations. At $\rm T_{eff}= 29,000\;$K,
the rbf transitions play a much more important role as N\,{\sc iii}
is the dominant ionization stage of the nitrogen atom.

\begin{table}
\begin{center}
\centering{
\caption{multi-MULTI Analysis at $\rm T_{eff}$= 29000 K, $ \log g$=4.0, and $\xi_t= 5.0\;\rm km\,s^{-1}$
\label{multi-multi_29kK}}
{\small
\begin{tabular*}{0.5\textwidth}{@{}l @{}l @{}l @{}l @{}l @{}l @{}l l@{}}
\hline\hline
Transition & \multicolumn{5}{c}{Perturbed transition} & Type & $\%$\\
\hline
3995.0\,\AA &   9& $\rightarrow $& 16& \multicolumn{1}{l}{($ 2p \;3s \;^1P^o$} &  \multicolumn{1}{l}{-$ 2p \;3p \;^1D$)} & \multicolumn{1}{l}{rbb} &    57.66\\
&   16& $\rightarrow $& 94& \multicolumn{1}{l}{($ 2p \;3p \;^1D$} &  \multicolumn{1}{l}{-$ 2s^2\;2p\;^2P^o$ N\,{\sc iii})} & \multicolumn{1}{l}{rbf} &     7.95\\
&    1& $\rightarrow $& 94& \multicolumn{1}{l}{($ 2p^2 \;^3P$} &  \multicolumn{1}{l}{-$ 2s^2\;2p\;^2P^o$ N\,{\sc iii})} & \multicolumn{1}{l}{rbf} &     6.83\\
&    2& $\rightarrow $& 94& \multicolumn{1}{l}{($ 2p2\;0p \;^1D$} &  \multicolumn{1}{l}{-$ 2s^2\;2p\;^2P^o$ N\,{\sc iii})} & \multicolumn{1}{l}{rbf} &     5.62\\
&   9& $\rightarrow $& 11& \multicolumn{1}{l}{($ 2p \;3s \;^1P^o$} &  \multicolumn{1}{l}{-$ 2p \;3p \;^1P$)} & \multicolumn{1}{l}{rbb} &     5.34\\
&   9& $\rightarrow $& 12& \multicolumn{1}{l}{($ 2p \;3s \;^1P^o$} &  \multicolumn{1}{l}{-$ 2p \;3p \;^3D$)} & \multicolumn{1}{l}{rbb} &     4.97\\
&   11& $\rightarrow $& 19& \multicolumn{1}{l}{($ 2p \;3p \;^1P$} &  \multicolumn{1}{l}{-$ 2p \;3d \;^1D^o$)}  & \multicolumn{1}{l}{rbb} &     4.97\\
&   4& $\rightarrow $& 94& \multicolumn{1}{l}{($2p^3 \;^5S^o$} &  \multicolumn{1}{l}{-$ 2s^2\;2p\;^2P^o$ N\,{\sc iii})} & \multicolumn{1}{l}{rbf} &     4.96\\
&   8& $\rightarrow $& 16& \multicolumn{1}{l}{($ 2p \;3s \;^3P^o$} &  \multicolumn{1}{l}{-$ 2p \;3p \;^1D$)} & \multicolumn{1}{l}{rbb} &     4.96\\
&   9& $\rightarrow $& 15& \multicolumn{1}{l}{($ 2p \;3s \;^1P^o$} &  \multicolumn{1}{l}{-$ 2p \;3p \;^3P$)} & \multicolumn{1}{l}{rbb} &     4.80\\ \cline{2-8}
6482.0\,\AA  &   9& $\rightarrow $& 11& \multicolumn{1}{l}{($ 2p \;3s \;^1P^o$} &  \multicolumn{1}{l}{-$ 2p \;3p \;^1P$)} & \multicolumn{1}{l}{rbb} &   125.08\\
&   11& $\rightarrow $& 94& \multicolumn{1}{l}{($ 2p \;3p \;^1P$} &  \multicolumn{1}{l}{-$ 2s^2\;2p\;^2P^o$ N\,{\sc iii})} & \multicolumn{1}{l}{rbf} &    24.15\\
&   9& $\rightarrow $& 16& \multicolumn{1}{l}{($ 2p \;3s \;^1P^o$} &  \multicolumn{1}{l}{-$ 2p \;3p \;^1D$)} & \multicolumn{1}{l}{rbb} &    18.65\\
&    1& $\rightarrow $& 94& \multicolumn{1}{l}{($ 2p^2 \;^3P$} &  \multicolumn{1}{l}{-$ 2s^2\;2p\;^2P^o$ N\,{\sc iii})} & \multicolumn{1}{l}{rbf} &    16.16\\
&  19& $\rightarrow $& 94& \multicolumn{1}{l}{($ 2p \;3d \;^1D^o$} &  \multicolumn{1}{l}{-$ 2s^2\;2p\;^2P^o$ N\,{\sc iii})} & \multicolumn{1}{l}{rbf} &    14.79\\
&   7& $\rightarrow $& 11& \multicolumn{1}{l}{($2p^3\;^1D^o$} &  \multicolumn{1}{l}{-$ 2p \;3p \;^1P$)} & \multicolumn{1}{l}{rbb} &    14.41\\
&    2& $\rightarrow $& 94& \multicolumn{1}{l}{($ 2p^2\;^1D$} &  \multicolumn{1}{l}{-$ 2s^2\;2p\;^2P^o$ N\,{\sc iii})} & \multicolumn{1}{l}{rbf} &    12.94\\
&   9& $\rightarrow $& 12& \multicolumn{1}{l}{($ 2p \;3s \;^1P^o$} &  \multicolumn{1}{l}{-$ 2p \;3p \;^3D$)} & \multicolumn{1}{l}{rbb} &    12.15\\
&   8& $\rightarrow $& 11& \multicolumn{1}{l}{($ 2p \;3s \;^3P^o$} &  \multicolumn{1}{l}{-$ 2p \;3p \;^1P$)} & \multicolumn{1}{l}{rbb} &    11.44\\
&   9& $\rightarrow $& 15& \multicolumn{1}{l}{($ 2p \;3s \;^1P^o$} &  \multicolumn{1}{l}{-$ 2p \;3p \;^3P$)} & \multicolumn{1}{l}{rbb} &    11.37\\
\hline
\end{tabular*}}}
\end{center}
\end{table}

\subsection{A Monte-MULTI Analysis}
\label{random-errors}

The accuracy of a non-LTE line formation calculation depends on many
factors, and an important one is the accuracy and completeness of the
basic atomic data used.  The inclusion of atomic data from many different
sources, all with different accuracies, represents a source of random
errors in the estimated equivalent widths. In order to quantify these
errors, a series of Monte Carlo simulations were carried out following
the procedure developed by \citet{sig96}. Two hundred random realizations
of the nitrogen atom were generated with the atomic data varied within
the bounds given in Table~\ref{atomic_data_variation}. The choices for
the adopted uncertainties in the atomic data are justified as follows:

\begin{itemize}

\item The oscillator strengths ($\it f$-values) of the bound-bound
radiative transitions were varied within $\pm$10 \% for ${\it f}\geq 0.1$
and $\pm$50 \% for weaker transitions.  These values were chosen because the
Opacity Project length and velocity $\it f$-values differ by approximately
these ranges. The difference between these two equivalent formalisms,
length and velocity, is a measure of the accuracy of the calculation
\citep{fro00}.

\item Stark damping widths were allowed to vary within $\pm$40\,\%, which
is on the order of the difference between our calculated Stark widths
using the OP formalism of \citet{sea88} and the available experimental
values-- see Table~\ref{gamma_ratios}.

\item The photoionization cross sections were allowed to vary within
$\pm$20$\%$. OP photoionization cross sections have uncertainties of
$\approx\,$10\,\% \citep{YS87} but this range was doubled to account
for possible errors in the photoionizing radiation field
predicted by the LTE, line-blanketed model atmospheres and used in the
calculation of the (fixed) photoionization rates.

\item Thermally-averaged collision strengths were allowed to vary over a
range determined by their source. The $\cal R$-matrix method represents
an accurate way to compute thermally-averaged collisional strengths
at low temperatures for low-lying levels in an ion.  \citet{hud04}
show that their results agree with the the results of previous {\cal
R}-matrix calculations within $\rm \approx\,10\,\%$, which we adopt
as the uncertainty in such collision strengths.  For majority of
radiatively-allowed transitions without ${\cal R}$-matrix collision
strengths, the impact parameter method \cite{sea62} was used, and a
factor of two uncertainly was employed. Note that \citet{sig96b} compared
${\cal R}$-matrix collision strengths to impact parameter predictions in
the case of Mg\,{\sc ii} and found about a factor of 2 in accuracy.

\item Collisional ionization rates are highly uncertain and the very crude
approximation of \citet{sea62} was used for all rates. An uncertainty of
a factor of 5 was assigned to all such values.

\end{itemize}

\begin{table}
\centering{
\caption{Rates of variation of the atomic data
\label{atomic_data_variation}}}
\begin{center}
{\small
\begin{tabular*}{0.43\textwidth}{@{}l @{}l }
\hline\hline
\multicolumn{1}{l}{Atomic parameter} & \multicolumn{1}{l}{Uncertainty}\\
\hline 
\multicolumn{1}{l}{$\it f$-value}& \multicolumn{1}{l}{$\pm$ 10\%  $\it f$-value $\ge$ 0.1}\\ 
& \multicolumn{1}{l}{$\pm$ 50\%   $\it f$-value $<$0.1}\\
\multicolumn{1}{l}{Stark widths} & \multicolumn{1}{l}{$\pm$ 40\%}\\
\multicolumn{1}{l}{Photoionization cross section} & \multicolumn{1}{l}{$\pm$ 20\%}\\
\multicolumn{1}{l}{Collision strength (Excitation)} & \\
\multicolumn{1}{l}{\hspace{0.5cm}{\cal R}-Matrix} & \multicolumn{1}{l}{$\pm$ 10\%}\\
\multicolumn{1}{l}{\hspace{0.5cm}Impact parameter} & \multicolumn{1}{l}{Factor of 2}\\
\multicolumn{1}{l}{Collisional strength (Ionization)}& \multicolumn{1}{l}{Factor of 5}\\
\hline
\end{tabular*}}
\end{center}
\vspace{0.1in}
\end{table}

\begin{figure}
\includegraphics[scale= 0.50]{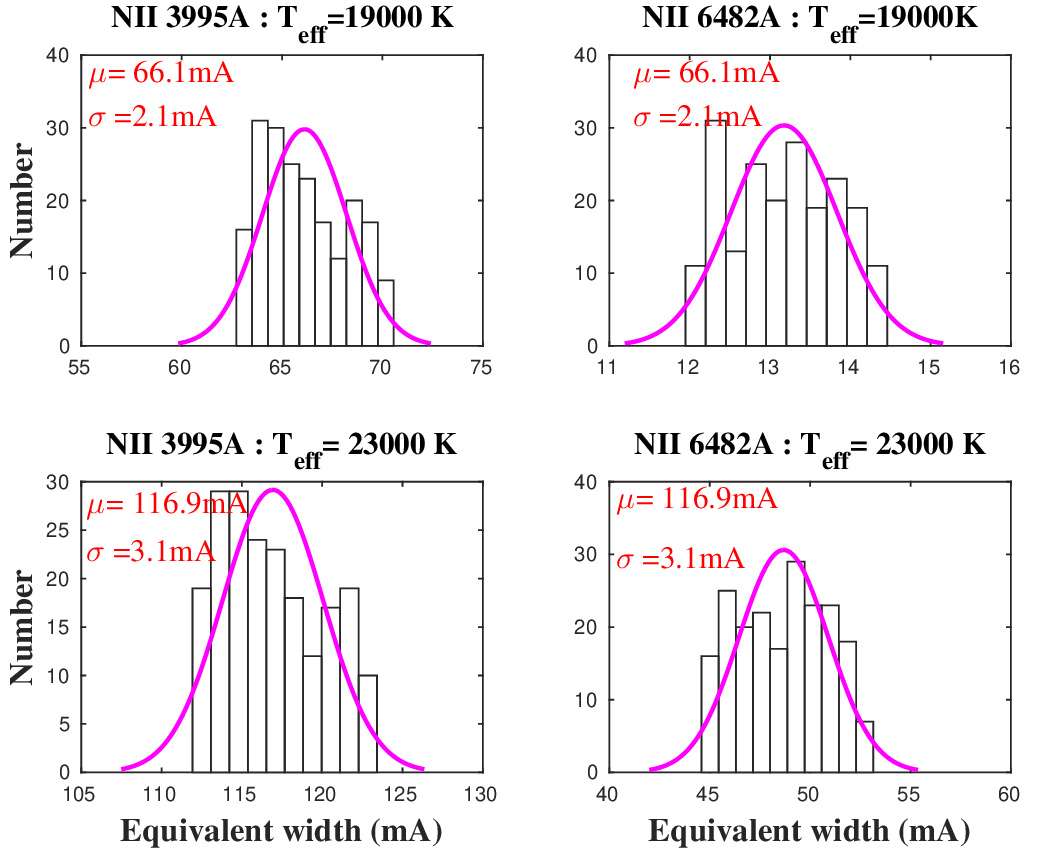}
\caption{The equivalent width distribution for the 200 N\,{\sc ii} model
atoms for $\lambda\,6482$ and
$\lambda\,3995$ at $\rm T_{eff}$= 19000 and 23000 K, $ \log g$,
equal to 4.0, $\xi_t= 5\;\rm km\,s^{-1}$, and solar nitrogen abundance were assumed.}
\label{MC-hist}
\end{figure}

A converged non-LTE solution was found for each of the 200
randomly-realized atoms, and the distribution of the predicted equivalent
widths was taken to estimate the uncertainty. Example distributions for
$\lambda\,3995$ and $\lambda\,6482$ are shown in Figure~\ref{MC-hist}
for the models with $T_{\rm eff}=19,000$ and 23,000\;K, with $\log g=
4.0$, $\xi_t= 5\;\rm km\,s^{-1}$ and the solar nitrogen abundance.
A Gaussian fit to each distribution gives the standard deviation
and hence the associated uncertainty due to inaccuracies in the
atomic data (taken to be $2\,\sigma$).  Tabular results are given
in Table~\ref{mc_eqw_res} for both transitions with $T_{\rm eff}$'s
between 15000 and 31000, and nitrogen abundances between 6.83 and 8.13.
Tables \ref{mc_eqw_res_g3.5_xi_t5} and \ref{mc_eqw_res_g4.5_xi_t5}
in appendix \ref{MCS_app} show the results of Monte Carlo simulations
for $\log g=\,$3.5 and 4.5, and $\xi=\,5.0\;\rm km\,s^{-1}$, over the
range of $T_{\rm eff}$ considered; given in each of these tables is the
average equivalent width of each transition and its $2\sigma$ variation.

In addition to the uncertainty itself, these random realizations allow
one to determine which individual rates most affect the uncertainty
in each transition's equivalent width.  This is different from the
previous multi-MULTI calculation as a realistic uncertainty for each
rate is used, as opposed to an arbitrary doubling. However, as the
individual rates are not varied one at a time, it is necessary to look
at the correlations between the equivalent width of the transition of
interest and the 200 scalings of each rate.  The largest correlation
coefficients of the N\;{\sc ii} $\lambda\,3995$ and $\lambda\,6482$
equivalent widths at $T_{\rm eff}$ of 19,000, 23,000 and 27,000 K,
$\log\,g=4.0$, $\xi_t=5\;\rm km\,s^{-1}$, and the solar nitrogen abundance
are shown in Table~(\ref{mc_corr}).  For 200 random realizations,
a correlation coefficient of 0.18 is a statistical significant at 1\%
level \citep{bev69}.  The variation in the predicted equivalent widths
of both transitions is most strongly correlated with the variations in
that transition's oscillator strength, as expected. The remainder of
the strongest correlations, at the level of $|r|\sim 0.22$, are with
collisional bound-bound rates between higher excitation levels. This
reflects the large uncertainties assigned to these rates as compared
to the oscillator strengths and $\cal R$-matrix collision strengths
(see Table~\ref{atomic_data_variation}).

\subsection{Limiting Accuracy of Nitrogen Abundances}

\begin{figure}
\includegraphics[scale= 0.48]{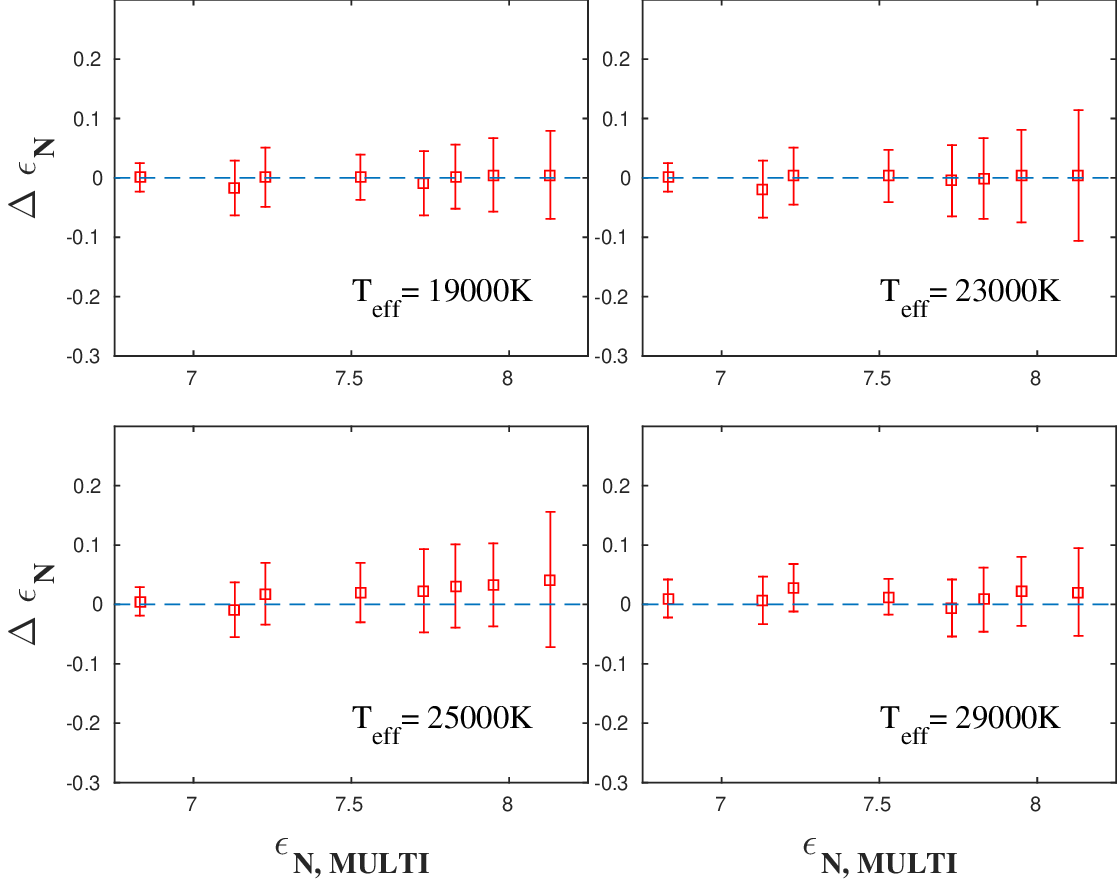}
\caption{Nitrogen abundances determined from the curves-of-growth
constructed from the 200 randomly realized atoms using the equivalent
widths of $\lambda\,3995$, $\lambda\,4447$ and $\lambda\,6482$ as the 
input observed values. Four $T_{\rm eff}$ are shown
and results were determined for a range of nitrogen abundances. The
error bars represent the uncertainties of the estimated abundances due
to inaccuracies in the atomic data at $2\,\sigma$. The blue dashed line
represents a difference of zero.}
\label{abund-hist}
\end{figure}

Finally, in order to quantify the ultimate accuracy of determined nitrogen
abundances due to uncertainties in the atomic data, the equivalent
widths predicted by the unperturbed atom for three singlet lines,
$\lambda\,3995$, $\lambda\,4447$ and $\lambda\,6482$, were used
as reference ``observed" equivalent widths for each $T_{\rm eff}$ assuming
$\log\,g=4.0$ and $\xi_t=5\;\rm km\,s^{-1}$. Then the curves of growth for
each of the 200 randomly-realized atoms were used to derive a nitrogen abundance
based on exactly the same stellar parameters and microturbulent value,
i.e.\ the ideal case. The dispersion in the abundances obtained from
the 200 curves-of-growth can then be taken as the limiting uncertainty
in the derived nitrogen abundance due to atomic data limitations. This process was then repeated for
all nitrogen abundances in the range considered, $\epsilon_{\rm N}=6.58$ to 8.53.
The results for four $T_{\rm eff}$ are shown in Figure~\ref{abund-hist}.

The figure shows that abundances obtained using the results of the
Monte Carlo simulations match the original abundances, and that the
errors in the estimated nitrogen abundances due to uncertain atomic
data increase with nitrogen abundance. For example, at $\rm T_{eff}=$
23,000 K, the uncertainty is $\rm \delta \epsilon=\pm 0.02\,$dex for $\rm
\epsilon_N=6.83\,$dex which rises to $\rm \delta \epsilon=\pm 0.11\,$dex
for $\rm \epsilon_N=8.13$\,dex. 

Figure~\ref{abund-hist} also shows that the estimated errors also
vary with  $T_{\rm eff}$.  At the same nitrogen abundance, such as
$\epsilon_N=7.83\,$dex, the abundance uncertainty is $\rm \delta
\epsilon=\pm 0.05\,$dex for $\rm T_{eff}= 19,000\;$K, $\rm \delta
\epsilon=\pm 0.07\,$dex for $\rm T_{eff}= 23,000\;$K and $\rm \delta
\epsilon=\pm 0.05\,$dex for $\rm T_{eff}= 29,000\;$K. The highest error
occurs for $\rm T_{eff}$ between 23,000~K and 25,000~K. In addition,
including uncertainty in $\rm T_{eff}$ of $\pm 1000.0\,$K causes
additional uncertainty in the estimated abundance by a factor of up to
$\approx\pm0.1\,$dex at the lowest temperatures, while uncertainty
in $\log g$ by $\pm0.25\,$dex adds additional uncertainty up to
$\pm0.05\,$dex in the estimated uncertainty of abundance.

\begin{table}
\caption{ Results of Monte Carlo Simulations for N\,{\sc ii} at $ \log g$=4.0, $\xi_t= 5.0\;\rm km\,s^{-1}$, and solar nitrogen abundance ($\rm \epsilon_{N,solar}= 7.83$): Correlation Coefficients
\label{mc_corr}}
\begin{center}
{\scriptsize
\begin{tabular*}{0.51\textwidth}{l@{\hskip 0.09in}c@{\hskip 0.09in}c@{\hskip -0.01in}c@{\hskip  -0.01in}c@{\hskip  -0.05in}c@{\hskip  -0.5in}c@{\hskip  -0.2in}c@{\hskip  0.2in}c@{\hskip  0.2in}c}
\hline\hline
\multicolumn{1}{c}{$\lambda$} & $\rm T_{eff}$ &  \multicolumn{5}{c}{Perturbed Transition} & \multicolumn{1}{l}{Type} & \multicolumn{1}{l}{r} \\  
\multicolumn{1}{c}{(\AA)} & (K) &   &\\
\hline
3995&19000&  9& $\rightarrow $& 16&  \multicolumn{1}{l}{($ 2p \;3s \;^1P^o$} & \multicolumn{1}{l}{-$ 2p \;3p \;^1D$)} & \multicolumn{1}{l}{rbb}&   0.988\\
& &  37& $\rightarrow $& 55& \multicolumn{1}{l}{($ 2p \;4f \;^1F$} &  \multicolumn{1}{l}{-$ 2p \;5d \;^1D^o$)}  & \multicolumn{1}{l}{cbb}&   -0.271\\
& &  28& $\rightarrow $& 35& \multicolumn{1}{l}{($ 2p \;4p \;^3P$} &  \multicolumn{1}{l}{-$ 2p \;4d \;^3D^o$)}  & \multicolumn{1}{l}{cbb}&    0.235\\
& &   9& $\rightarrow $& 16& \multicolumn{1}{l}{($ 2p \;3s \;^1P^o$} &  \multicolumn{1}{l}{-$ 2p \;3p \;^1D$)} & \multicolumn{1}{l}{stk}&    0.222\\
& &  94& $\rightarrow $&104& \multicolumn{1}{l}{($ 2s^2\;2p\;^2P^o$ N\,{\sc iii}} &  \multicolumn{1}{l}{-$ 2s^2\;3d\;^2D$ N\,{\sc iii})} & \multicolumn{1}{l}{cbb}&    0.221\\\cline{4-8}
&23000&  9& $\rightarrow $& 16&  \multicolumn{1}{l}{($ 2p \;3s \;^1P^o$} & \multicolumn{1}{l}{-$ 2p \;3p \;^1D$)} & \multicolumn{1}{l}{rbb}&   0.983\\
& &  37& $\rightarrow $& 55& \multicolumn{1}{l}{($ 2p \;4f \;^1F$} &  \multicolumn{1}{l}{-$ 2p \;5d \;^1D^o$)}  & \multicolumn{1}{l}{cbb}&   -0.270\\
& &  28& $\rightarrow $& 35& \multicolumn{1}{l}{($ 2p \;4p \;^3P$} &  \multicolumn{1}{l}{-$ 2p \;4d \;^3D^o$)}  & \multicolumn{1}{l}{cbb}&    0.234\\
& &  94& $\rightarrow $&104& \multicolumn{1}{l}{($ 2s^2\;2p\;^2P^o$ N\,{\sc iii}} &  \multicolumn{1}{l}{-$ 2s^2\;3d\;^2D$ N\,{\sc iii})} & \multicolumn{1}{l}{cbb}&    0.228\\
& &   8& $\rightarrow $& 49& \multicolumn{1}{l}{($ 2p \;3s \;^3P^o$} &  \multicolumn{1}{l}{-$ 2p \;5p \;^3D$)} & \multicolumn{1}{l}{cbb}&   -0.223\\\cline{3-8}
&27000&  9& $\rightarrow $& 16&  \multicolumn{1}{l}{($ 2p \;3s \;^1P^o$} & \multicolumn{1}{l}{-$ 2p \;3p \;^1D$)} & \multicolumn{1}{l}{rbb}&   0.967\\
& &  37& $\rightarrow $& 55& \multicolumn{1}{l}{($ 2p \;4f \;^1F$} &  \multicolumn{1}{l}{-$ 2p \;5d \;^1D^o$)}  & \multicolumn{1}{l}{cbb}&   -0.265\\
& &  94& $\rightarrow $&104& \multicolumn{1}{l}{($ 2s^2\;2p\;^2P^o$ N\,{\sc iii}} &  \multicolumn{1}{l}{-$ 2s^2\;3d\;^2D$ N\,{\sc iii})} & \multicolumn{1}{l}{cbb}&    0.234\\
& &   8& $\rightarrow $& 49& \multicolumn{1}{l}{($ 2p \;3s \;^3P^o$} &  \multicolumn{1}{l}{-$ 2p \;5p \;^3D$)} & \multicolumn{1}{l}{cbb}&   -0.231\\
& &  28& $\rightarrow $& 35& \multicolumn{1}{l}{($ 2p \;4p \;^3P$} &  \multicolumn{1}{l}{-$ 2p \;4d \;^3D^o$)}  & \multicolumn{1}{l}{cbb}&    0.225\\\cline{3-8}
6482&19000&  9& $\rightarrow $& 11&  \multicolumn{1}{l}{($ 2p \;3s \;^1P^o$} & \multicolumn{1}{l}{-$ 2p \;3p \;^1P$)} & \multicolumn{1}{l}{rbb}&   0.988\\
& &  14& $\rightarrow $& 73& \multicolumn{1}{l}{($ 2p \;3p \;^3S$} &  \multicolumn{1}{l}{-$ 2p \;6s \;^3P^o$)}  & \multicolumn{1}{l}{cbb}&   -0.244\\
& &  55& $\rightarrow $& 94& \multicolumn{1}{l}{($ 2p \;5d \;^1D^o$} &  \multicolumn{1}{l}{-$ 2s^2\;2p\;^2P^o$ N\,{\sc iii})} & \multicolumn{1}{l}{cbf}&    0.227\\
& &  47& $\rightarrow $& 77& \multicolumn{1}{l}{($ 2p \;5s \;^1P^o$} &  \multicolumn{1}{l}{-$ 2p \;6p \;^1P$)} & \multicolumn{1}{l}{rbb}&   -0.219\\
& &  56& $\rightarrow $& 94& \multicolumn{1}{l}{($ 2p \;5d \;^3D^o$} &  \multicolumn{1}{l}{-$ 2s^2\;2p\;^2P^o$ N\,{\sc iii})} & \multicolumn{1}{l}{cbf}&   -0.218\\\cline{3-8.5}
&23000&  9& $\rightarrow $& 11&  \multicolumn{1}{l}{($ 2p \;3s \;^1P^o$} & \multicolumn{1}{l}{-$ 2p \;3p \;^1P$)} & \multicolumn{1}{l}{rbb}&   0.969\\
& &  14& $\rightarrow $& 73& \multicolumn{1}{l}{($ 2p \;3p \;^3S$} &  \multicolumn{1}{l}{-$ 2p \;6s \;^3P^o$)}  & \multicolumn{1}{l}{cbb}&   -0.251\\
& &  55& $\rightarrow $& 94& \multicolumn{1}{l}{($ 2p \;5d \;^1D^o$} &  \multicolumn{1}{l}{-$ 2s^2\;2p\;^2P^o$ N\,{\sc iii})} & \multicolumn{1}{l}{cbf}&    0.231\\
& &  47& $\rightarrow $& 77& \multicolumn{1}{l}{($ 2p \;5s \;^1P^o$} &  \multicolumn{1}{l}{-$ 2p \;6p \;^1P$)} & \multicolumn{1}{l}{rbb}&   -0.210\\
& &  56& $\rightarrow $& 80& \multicolumn{1}{l}{($ 2p \;5d \;^3D^o$} &  \multicolumn{1}{l}{-$ 2p \;6p \;^3P$)} & \multicolumn{1}{l}{stk}&   -0.210\\\cline{3-8}
&27000&  9& $\rightarrow $& 11&  \multicolumn{1}{l}{($ 2p \;3s \;^1P^o$} & \multicolumn{1}{l}{-$ 2p \;3p \;^1P$)} & \multicolumn{1}{l}{rbb}&   0.910\\
& &  14& $\rightarrow $& 73& \multicolumn{1}{l}{($ 2p \;3p \;^3S$} &  \multicolumn{1}{l}{-$ 2p \;6s \;^3P^o$)}  & \multicolumn{1}{l}{cbb}&   -0.252\\
& &  55& $\rightarrow $& 94& \multicolumn{1}{l}{($ 2p \;5d \;^1D^o$} &  \multicolumn{1}{l}{-$ 2s^2\;2p\;^2P^o$ N\,{\sc iii})} & \multicolumn{1}{l}{cbf}&    0.225\\
& &  20& $\rightarrow $& 43& \multicolumn{1}{l}{($ 2p \;3d \;^3D^o$} &  \multicolumn{1}{l}{-$ 2p \;4f \;^3D$)} & \multicolumn{1}{l}{rbb}&   -0.216\\
& &  56& $\rightarrow $& 80& \multicolumn{1}{l}{($ 2p \;5d \;^3D^o$} &  \multicolumn{1}{l}{-$ 2p \;6p \;^3P$)} & \multicolumn{1}{l}{stk}&   -0.213\\
\hline
\hline
\end{tabular*}}
\end{center}
\noindent{Notes: All correlation coefficients are statistically significant to less than 1\% level, and "stk" refers to variation of the Stark width of the radiative transition.} 
\end{table}

\subsection{Possible Systematic Errors}

Systematic errors are more difficult to quantify than the random errors
in the atomic data. Possible systematic errors can originate from many
sources, such the use of LTE stellar atmospheres with fixed elemental
abundances, fixed photoionization rates, and a truncated atomic model
in terms of included energy levels.

Widely available non-LTE stellar atmosphere models do not readily provide
grids of $J_\nu$ as a function of optical depth needed for calculating
the photoionization rates. On the other side, the opacity distribution
function treatment of \citet{kur79} provides $J_\nu$  over manageable
grids; this issue is discussed by \citet{prz01}.

Another source of systematic error is the completeness of the nitrogen
atom. Using atomic models with only a few energy levels can result
in large non-LTE effects, particularly for trace ions with low-lying
levels with photoionization thresholds in the short-wavelength region
of the Balmer continuum. Such calculations can predict too much
overionization in small atomic models because collisional coupling to
the dominant continuum is artificially suppressed \citep{sig96}.  On the
other hand, there is a physical limit to how many energy levels can be
straightforwardly added to a non-LTE calculation; highly excited levels
are only weakly coupled to their parent ion and can be disrupted by the
surrounding plasma. This is most naturally described by introducing an
occupation probability, between 0 and 1, for each bound level to exist
\citep{hum88} and reformulating the statistical equilibrium equations
to include the occupation probabilities \citep{hub94}.

\begin{figure}
\centering
\includegraphics[scale= 0.4]{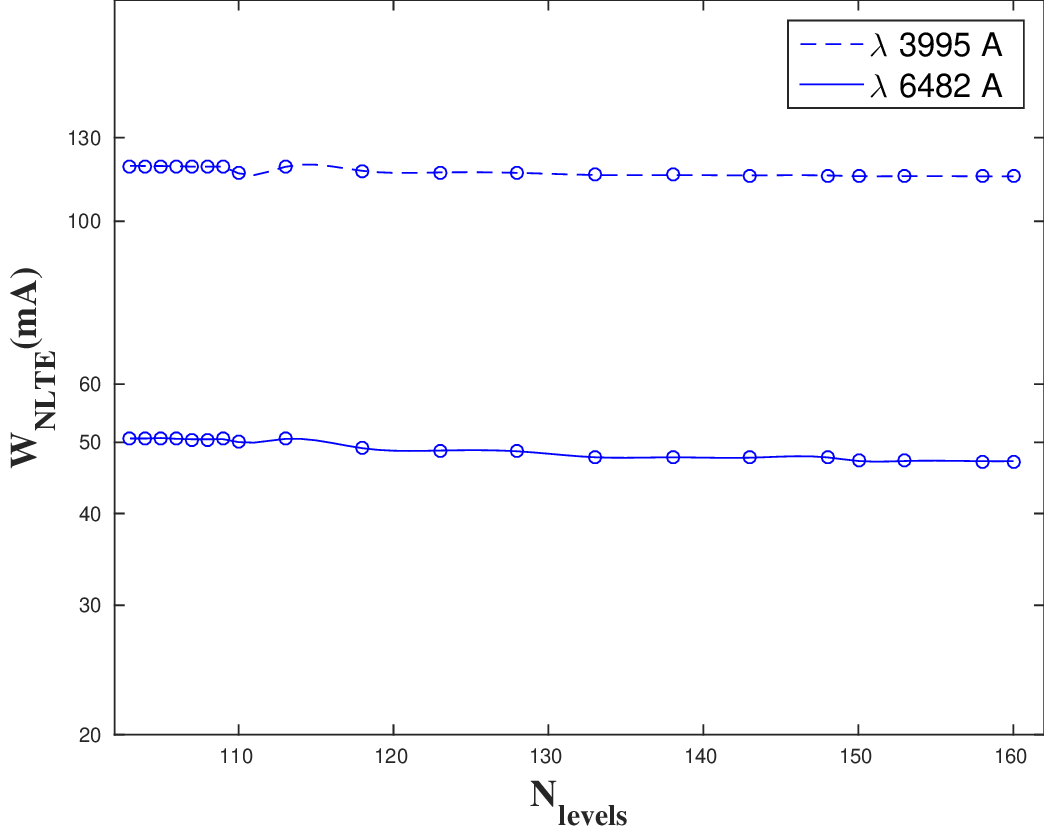}
\caption{The change in the predicted equivalent widths of $\lambda\,3995$
and $\lambda\,6482$ with an increasing number of levels
in the nitrogen atom at $\rm T_{eff}=$ 23,000 K, $\log g=$ 4.0, $\xi_t=
5\;\rm km\,s^{-1}$, and the solar nitrogen abundance.}
\label{eqw_nlevels}
\end{figure}

In order to test our N\,{\sc ii} atom for completeness, we have
systematically increased the number of included N\,{\sc ii} energy
levels from 93 to 150, and for each increase, recomputed the non-LTE
solution. Figure~\ref{eqw_nlevels} shows the effect on the equivalent
widths of $\lambda\,3995$ and $\lambda\,6482$.  The effect is quite small,
a few percent at most, and hence we do not consider the size of the
N\,{\sc ii} atom used in the present work to be a significant source of
uncertainty for the transitions of focus. This lack of strong dependence
on the number of non-LTE levels is consistent with the nature of the
non-LTE effects in N\,{\sc ii}. Because the photoionization thresholds
of low-lying N\,{\sc ii} levels are in the Lyman continuum, there is
little predicted non-LTE overionization in the line forming region
and hence collisional coupling to the parent N\,{\sc iii} continuum is
less important.

\section{Discussion and Conclusion}
\label{disc}

We have presented new, non-LTE line calculations for N\;{\sc ii}, using
the MULTI code of \citet{car92}, over the range of stellar $T_{\rm eff}$,
$\log\,g$, and microturbulent velocities appropriate to the main sequence
B~stars.  Grids of equivalent widths for commonly-used, strong N\;{\sc
ii} lines in elemental abundance analysis have been provided.  A detailed
error analysis was performed, and the error bounds on the tabulated
equivalent widths due to atomic data uncertainties provided.

We find reasonable agreement with the most recently, tabulated non-LTE
equivalents widths, those of \citet{kor99}, but, in general, we find weaker
non-LTE effects. We attribute this to the much larger number of radiative
bound-bound transitions explicitly included in the non-LTE calculation, as
opposed to treating many weaker transitions with fixed rates. In addition,
our careful treatment of quadratic Stark broadening makes our N\,{\sc ii}
equivalent widths more reliable at their maximum strengths.

We systematically investigated the completeness of our atomic model in terms
of included energy levels. In addition, an extensive Monte-MULTI calculation
quantified the effect of atomic data limitations on the N\,{\sc ii}
equivalent width predictions. In general, near their peak strengths, we find
that the limiting accuracy of using non-LTE equivalents widths computed with
existing atomic data is $\approx\pm0.1\;$dex, although this range is somewhat
larger for the highest $T_{\rm eff}$ considered.

To be most applicable to main sequence B stars, the effect of
gravitational darkening, particularly the latitude-dependent $T_{\rm
eff}$, on the optical N\,{\sc ii} spectrum should be investigated, and for
the Be stars, the potential of disk emission should be quantified. These
effects will be investigated in a subsequent work.

\vspace{0.3in}
This work is supported by the Canadian Natural Sciences and Engineering
Research Council (NESRC) through a Discovery Grant to TAAS.







\appendix

\section{Equivalent Widths}
\label{app_EW}

This appendix contains the non-LTE equivalent widths (in m\AA) for
N\,{\sc ii} transitions $\lambda\,3995$ and $\lambda\,6482$ as a
function of $T_{\rm eff}$, $\log\,g$, and nitrogen abundance, all for a
microturbulence of $\xi_t =5\;\rm km\,s^{-1}$. The tabulated
widths are those of the unperturbed atom. Equivalent widths
for the other transitions in Table~\ref{rbb-trans-data}, and for other
$\xi_t$ values, are given in the on-line tables.

\newpage
\begin{landscape}
\begin{table}
\centering{
\caption{Non-LTE equivalent widths (in m\AA) for N\;{\sc ii} $\lambda$ 3995\,\AA\;at $\xi_t =5\;\rm km\,s^{-1}$.
These equivalent widths are for the unperturbed reference atom.
\label{multi_wv3995_vturb5_2}}}
\begin{center}
{\small
\begin{tabular}{c rrrrrrrrrrrrrrrrrrrrrrr}
\hline\hline
\multicolumn{1}{l}{$\rm T_{eff}(\rm K)$, $\log g$}&\multicolumn{14}{c}{$\rm\Delta\,\epsilon_N=\epsilon_{N}-7.83$}\\
& \multicolumn{2}{c}{-1.00}& \multicolumn{2}{c}{-0.70}& \multicolumn{2}{c}{-0.60}& \multicolumn{2}{c}{-0.30}& \multicolumn{2}{c}{0.00}& \multicolumn{2}{c}{+0.18}& \multicolumn{2}{c}{+0.30}\\
& $W_\lambda$ & $W/W_*$& $W_\lambda$ & $W/W_*$& $W_\lambda$ & $W/W_*$& $W_\lambda$ & $W/W_*$& $W_\lambda$ & $W/W_*$& $W_\lambda$ & $W/W_*$& $W_\lambda$ & $W/W_*$\\
\hline
15000 ,  3.5 & 6.4 & 1.11 &11.2 & 1.13 &13.2 & 1.14 &21.2 & 1.16 &32.1 & 1.17 &40.1 & 1.18 &46.4 & 1.18 \\
17000 ,  3.5 &14.9 & 1.14 &24.4 & 1.16 &28.2 & 1.17 &42.3 & 1.18 &60.0 & 1.19 &72.1 & 1.20 &81.4 & 1.20 \\
19000 ,  3.5 &26.1 & 1.15 &41.6 & 1.17 &47.5 & 1.18 &68.3 & 1.20 &92.6 & 1.21 &108.4 & 1.22 &120.4 & 1.22 \\
21000 ,  3.5 &35.3 & 1.13 &56.4 & 1.17 &64.3 & 1.18 &91.3 & 1.21 &121.1 & 1.23 &139.9 & 1.24 &153.9 & 1.24 \\
23000 ,  3.5 &35.3 & 1.07 &58.4 & 1.11 &67.3 & 1.13 &98.0 & 1.18 &131.9 & 1.22 &152.7 & 1.24 &168.0 & 1.25 \\
25000 ,  3.5 &25.7 & 1.02 &45.1 & 1.06 &53.0 & 1.08 &82.0 & 1.13 &115.7 & 1.19 &136.7 & 1.21 &152.1 & 1.23 \\
27000 ,  3.5 &14.1 & 0.96 &27.5 & 1.03 &33.4 & 1.05 &57.0 & 1.11 &87.2 & 1.16 &106.7 & 1.19 &121.0 & 1.21 \\
29000 ,  3.5 & 3.1 & 0.50 & 8.8 & 0.73 &11.8 & 0.79 &25.5 & 0.93 &47.4 & 1.04 &63.7 & 1.09 &76.3 & 1.11 \\
31000 ,  3.5 &-1.1 & -0.66 &-1.3 & -0.41 &-1.3 & -0.30 & 0.4 & 0.05 & 6.4 & 0.37 &12.9 & 0.52 &19.3 & 0.61 \\
\hline
15000 ,  4.0 & 3.6 & 1.07 & 6.6 & 1.08 & 7.9 & 1.08 &13.3 & 1.10 &21.3 & 1.11 &27.3 & 1.12 &32.3 & 1.12 \\
17000 ,  4.0 & 9.0 & 1.08 &15.4 & 1.10 &18.1 & 1.11 &28.5 & 1.12 &42.4 & 1.13 &52.3 & 1.14 &60.1 & 1.14 \\
19000 ,  4.0 &17.3 & 1.09 &28.6 & 1.12 &33.0 & 1.12 &49.5 & 1.14 &69.9 & 1.15 &83.7 & 1.16 &94.2 & 1.16 \\
21000 ,  4.0 &26.7 & 1.09 &43.4 & 1.13 &49.8 & 1.14 &72.5 & 1.16 &98.8 & 1.17 &115.8 & 1.18 &128.6 & 1.18 \\
23000 ,  4.0 &32.4 & 1.07 &53.2 & 1.11 &61.2 & 1.12 &88.9 & 1.16 &119.7 & 1.19 &138.9 & 1.20 &153.2 & 1.20 \\
25000 ,  4.0 &29.3 & 1.01 &49.9 & 1.06 &58.1 & 1.08 &87.2 & 1.13 &119.9 & 1.17 &140.1 & 1.19 &154.9 & 1.20 \\
27000 ,  4.0 &20.7 & 0.97 &37.3 & 1.02 &44.3 & 1.04 &70.5 & 1.09 &101.9 & 1.14 &121.6 & 1.16 &136.0 & 1.17 \\
29000 ,  4.0 &11.2 & 0.86 &22.4 & 0.94 &27.5 & 0.96 &48.2 & 1.03 &75.7 & 1.08 &93.8 & 1.11 &107.2 & 1.13 \\
31000 ,  4.0 & 2.9 & 0.44 & 7.9 & 0.63 &10.4 & 0.68 &22.4 & 0.81 &41.7 & 0.91 &56.2 & 0.96 &67.5 & 0.99 \\
\hline
15000 ,  4.5 & 2.1 & 1.03 & 3.9 & 1.04 & 4.7 & 1.05 & 8.3 & 1.06 &14.0 & 1.07 &18.4 & 1.07 &22.2 & 1.08 \\
17000 ,  4.5 & 5.4 & 1.04 & 9.6 & 1.06 &11.4 & 1.06 &18.9 & 1.07 &29.5 & 1.08 &37.4 & 1.09 &43.7 & 1.09 \\
19000 ,  4.5 &11.1 & 1.05 &19.0 & 1.07 &22.2 & 1.08 &34.9 & 1.09 &51.5 & 1.10 &63.1 & 1.11 &72.3 & 1.11 \\
21000 ,  4.5 &18.6 & 1.06 &31.2 & 1.08 &36.2 & 1.09 &54.7 & 1.11 &77.3 & 1.12 &92.5 & 1.13 &104.1 & 1.13 \\
23000 ,  4.5 &25.7 & 1.05 &42.8 & 1.08 &49.5 & 1.09 &73.4 & 1.12 &101.1 & 1.14 &118.8 & 1.15 &132.2 & 1.15 \\
25000 ,  4.5 &27.9 & 1.02 &47.3 & 1.06 &54.9 & 1.07 &82.1 & 1.11 &112.8 & 1.14 &132.0 & 1.15 &146.3 & 1.15 \\
27000 ,  4.5 &23.0 & 0.97 &40.6 & 1.01 &47.8 & 1.02 &74.5 & 1.07 &105.5 & 1.11 &124.9 & 1.12 &139.1 & 1.13 \\
29000 ,  4.5 &15.3 & 0.91 &28.6 & 0.96 &34.4 & 0.98 &57.3 & 1.03 &86.0 & 1.07 &104.4 & 1.09 &117.8 & 1.10 \\
31000 ,  4.5 & 7.4 & 0.73 &15.7 & 0.83 &19.5 & 0.86 &36.1 & 0.93 &59.5 & 0.99 &75.6 & 1.02 &87.7 & 1.04 \\
\hline
\end{tabular}}
\end{center}
\end{table}
\end{landscape}
\clearpage

\begin{landscape}
\begin{table}
\centering{
\caption{Non-LTE equivalent widths (in m\AA) for N\;{\sc ii} $\lambda$ 6482 \,\AA\;at $\xi_t =5\;\rm km\,s^{-1}$.
These equivalent widths are for the unperturbed reference atom.
\label{multi_wv6482_zeta5}}}
\begin{center}
{\small
\begin{tabular}{c rrrrrrrrrrrrrrrrrrrrrrr}
\hline\hline
\multicolumn{1}{l}{$\rm T_{eff}(\rm K)$, $\log(g)$}&\multicolumn{14}{c}{$\rm\Delta\,\epsilon_N=\epsilon_{N}- 7.83$}\\
& \multicolumn{2}{c}{-1.00}& \multicolumn{2}{c}{-0.70}& \multicolumn{2}{c}{-0.60}& \multicolumn{2}{c}{-0.30}& \multicolumn{2}{c}{0.00}& \multicolumn{2}{c}{+0.18}& \multicolumn{2}{c}{+0.30}\\
& $W_\lambda$ & $W/W_*$& $W_\lambda$ & $W/W_*$& $W_\lambda$ & $W/W_*$& $W_\lambda$ & $W/W_*$& $W_\lambda$ & $W/W_*$& $W_\lambda$ & $W/W_*$& $W_\lambda$ & $W/W_*$\\
\hline
15000 ,  3.5 & 0.2 & 0.90 & 0.5 & 0.98 & 0.6 & 1.00 & 1.2 & 1.07 & 2.5 & 1.13 & 3.7 & 1.17 & 4.8 & 1.19 \\
17000 ,  3.5 & 0.9 & 0.96 & 1.9 & 1.05 & 2.4 & 1.08 & 4.8 & 1.15 & 9.0 & 1.21 &12.7 & 1.24 &16.1 & 1.26 \\
19000 ,  3.5 & 2.7 & 0.98 & 5.7 & 1.09 & 7.2 & 1.12 &13.8 & 1.21 &24.4 & 1.28 &32.9 & 1.31 &40.2 & 1.33 \\
21000 ,  3.5 & 5.8 & 0.96 &12.2 & 1.09 &15.3 & 1.14 &28.8 & 1.25 &49.2 & 1.34 &64.5 & 1.39 &76.9 & 1.42 \\
23000 ,  3.5 & 7.3 & 0.88 &16.0 & 1.03 &20.3 & 1.08 &39.6 & 1.24 &68.7 & 1.36 &89.9 & 1.43 &106.5 & 1.46 \\
25000 ,  3.5 & 5.0 & 0.74 &11.5 & 0.90 &14.9 & 0.95 &31.7 & 1.14 &59.8 & 1.30 &81.6 & 1.39 &99.0 & 1.44 \\
27000 ,  3.5 & 2.0 & 0.50 & 5.4 & 0.69 & 7.3 & 0.76 &17.4 & 0.97 &37.2 & 1.18 &54.6 & 1.29 &69.6 & 1.36 \\
29000 ,  3.5 &-1.3 & -0.70 &-0.6 & -0.17 &-0.1 & -0.01 & 3.9 & 0.42 &13.7 & 0.80 &24.0 & 0.98 &33.8 & 1.10 \\
31000 ,  3.5 &-3.2 & -5.00 &-5.0 & -3.84 &-5.5 & -3.42 &-6.7 & -2.08 &-6.1 & -0.94 &-3.8 & -0.40 &-0.7 & -0.06 \\
\hline
15000 ,  4.0 & 0.1 & 0.84 & 0.2 & 0.91 & 0.3 & 0.93 & 0.6 & 1.00 & 1.2 & 1.06 & 1.8 & 1.09 & 2.4 & 1.11 \\
17000 ,  4.0 & 0.4 & 0.91 & 0.9 & 0.98 & 1.1 & 1.01 & 2.3 & 1.08 & 4.6 & 1.14 & 6.7 & 1.16 & 8.6 & 1.18 \\
19000 ,  4.0 & 1.3 & 0.95 & 2.9 & 1.04 & 3.6 & 1.07 & 7.2 & 1.14 &13.3 & 1.20 &18.6 & 1.23 &23.2 & 1.24 \\
21000 ,  4.0 & 3.3 & 0.95 & 7.0 & 1.06 & 8.8 & 1.10 &16.9 & 1.19 &30.0 & 1.25 &40.3 & 1.28 &48.9 & 1.30 \\
23000 ,  4.0 & 5.6 & 0.91 &12.1 & 1.04 &15.2 & 1.09 &29.2 & 1.20 &50.6 & 1.29 &66.5 & 1.33 &79.2 & 1.35 \\
25000 ,  4.0 & 5.6 & 0.81 &12.5 & 0.95 &16.0 & 1.00 &32.1 & 1.15 &57.4 & 1.26 &76.3 & 1.32 &91.3 & 1.35 \\
27000 ,  4.0 & 3.6 & 0.68 & 8.3 & 0.83 &10.8 & 0.88 &23.3 & 1.04 &45.2 & 1.19 &62.9 & 1.26 &77.3 & 1.31 \\
29000 ,  4.0 & 1.4 & 0.43 & 4.0 & 0.62 & 5.4 & 0.69 &13.0 & 0.88 &28.2 & 1.07 &41.8 & 1.16 &53.7 & 1.22 \\
31000 ,  4.0 &-0.9 & -0.50 &-0.1 & -0.04 & 0.4 & 0.10 & 3.8 & 0.46 &11.8 & 0.77 &20.0 & 0.92 &27.8 & 1.01 \\
\hline
15000 ,  4.5 & 0.1 & 0.81 & 0.1 & 0.86 & 0.1 & 0.88 & 0.3 & 0.94 & 0.6 & 0.99 & 0.9 & 1.02 & 1.2 & 1.04 \\
17000 ,  4.5 & 0.2 & 0.86 & 0.4 & 0.93 & 0.6 & 0.95 & 1.2 & 1.01 & 2.4 & 1.06 & 3.5 & 1.09 & 4.6 & 1.11 \\
19000 ,  4.5 & 0.7 & 0.91 & 1.4 & 0.98 & 1.8 & 1.01 & 3.7 & 1.07 & 7.2 & 1.12 &10.3 & 1.15 &13.2 & 1.16 \\
21000 ,  4.5 & 1.8 & 0.93 & 3.8 & 1.02 & 4.8 & 1.05 & 9.4 & 1.12 &17.4 & 1.17 &24.1 & 1.20 &29.9 & 1.21 \\
23000 ,  4.5 & 3.6 & 0.92 & 7.7 & 1.03 & 9.7 & 1.06 &18.8 & 1.15 &33.3 & 1.21 &44.6 & 1.23 &53.9 & 1.25 \\
25000 ,  4.5 & 4.9 & 0.86 &10.7 & 0.98 &13.5 & 1.02 &26.5 & 1.13 &46.7 & 1.20 &61.9 & 1.23 &74.1 & 1.25 \\
27000 ,  4.5 & 4.2 & 0.75 & 9.3 & 0.88 &11.9 & 0.92 &24.5 & 1.05 &45.1 & 1.15 &61.1 & 1.19 &74.1 & 1.22 \\
29000 ,  4.5 & 2.5 & 0.63 & 5.8 & 0.76 & 7.6 & 0.81 &16.6 & 0.95 &33.0 & 1.08 &46.7 & 1.14 &58.3 & 1.17 \\
31000 ,  4.5 & 0.7 & 0.30 & 2.4 & 0.53 & 3.4 & 0.59 & 8.6 & 0.79 &19.2 & 0.96 &28.9 & 1.04 &37.7 & 1.09 \\
\hline
\end{tabular}}
\end{center}
\end{table}
\end{landscape}

\section{Monte-Carlo Simulations}\label{MCS_app}

This appendix contains the results of the Monte Carlo simulations for
N\,{\sc ii} transitions $\lambda\,3995$ and $\lambda\,6482$ as a
function of $T_{\rm eff}$, $\log\,g$ and the nitrogen abundance, all for
$\xi_t= 5\,\rm km\,s^{-1}$. Tabulated are the average equivalent width
(in m\AA) and its standard deviation over the 200 random realizations
of the nitrogen atomic model. There are small differences between the
average equivalent widths tabulated here and the widths predicted by
the unperturbed atom, given previously in Appendix~A. The atomic data
scalings of Table~\ref{atomic_data_variation} for the less certain rates
(non-$\cal R$-matrix collisional rates, for example) are not symmetric
about the default rate and this can lead to differences between the
predictions of the reference atom and the ensemble average.

\clearpage
\begin{table}
\centering{
\caption{ Results of Monte Carlo Simulations for N\,{\sc ii} at $\log g$=3.5, and $\xi_t= 5.0\;\rm km\,s^{-1}$: average equivalent widths, $<W_{\lambda}>$, and the expected error, $\sigma$
\label{mc_eqw_res_g3.5_xi_t5}}}
\begin{center}
{\scriptsize
\begin{tabular*}{0.51\textwidth}{l@{\hskip 0.09in}c@{\hskip 0.09in}c@{\hskip -0.05in}c@{\hskip  -0.05in}c@{\hskip  0.09in}c@{\hskip  0.09in}c@{\hskip  0.09in}c@{\hskip  -0.05in}c@{\hskip  -0.05in}c}
\hline\hline
\multicolumn{1}{c}{$\lambda$}& $\epsilon_N$ & $\rm T_{eff}$ & \multicolumn{1}{l}{\tiny{$<W_{\lambda}>$}} & \multicolumn{1}{c}{$2\sigma$} &\multicolumn{1}{c}{$\lambda$}& $\epsilon_N$ & $\rm T_{eff}$ & \multicolumn{1}{l}{\tiny{$<W_{\lambda}>$}} & \multicolumn{1}{l}{$2\,\sigma$}\\     
\multicolumn{1}{l}{(\AA)}&& (K)& \multicolumn{1}{l}{(m\AA)} & (m\AA) &\multicolumn{1}{l}{(\AA)}&& (K)& \multicolumn{1}{l}{(m\AA)} & (m\AA) \\
\hline
3995 & 6.830 & 15000.0 &  5.8 & 0.60 & 6482 & 6.83 & 15000.0 &  0.2 & 0.03 \\
&& 19000.0 & 25.3 & 2.36 &&& 19000.0 &  2.7 & 0.34 \\
&& 23000.0 & 36.3 & 3.65 &&& 23000.0 &  6.5 & 1.08 \\
&& 27000.0 & 14.8 & 1.94 &&& 27000.0 &  1.1 & 0.45 \\
&& 31000.0 & -1.4 & 0.20 &&& 31000.0 & -3.5 & 0.28 \\\cline{3-5}  \cline{8-10}
 & 7.230 & 15000.0 & 12.1 & 1.11 && 7.23 & 15000.0 &  0.6 & 0.07 \\
&& 19000.0 & 46.0 & 3.49 &&& 19000.0 &  7.0 & 0.81 \\
&& 23000.0 & 67.9 & 5.32 &&& 23000.0 & 18.3 & 2.52 \\
&& 27000.0 & 33.9 & 3.77 &&& 27000.0 &  4.8 & 1.24 \\
&& 31000.0 & -0.9 & 0.60 &&& 31000.0 & -6.3 & 0.49 \\\cline{3-5}  \cline{8-10}
 & 7.530 & 15000.0 & 19.6 & 1.62 && 7.53 & 15000.0 &  1.2 & 0.14 \\
&& 19000.0 & 66.1 & 4.29 &&& 19000.0 & 13.4 & 1.43 \\
&& 23000.0 & 97.7 & 6.16 &&& 23000.0 & 36.1 & 4.19 \\
&& 27000.0 & 57.3 & 5.27 &&& 27000.0 & 12.6 & 2.47 \\
&& 31000.0 &  2.7 & 1.38 &&& 31000.0 & -8.5 & 0.66 \\\cline{3-5}  \cline{8-10}
 & 7.830 & 15000.0 & 29.9 & 2.23 && 7.83 & 15000.0 &  2.5 & 0.27 \\
&& 19000.0 & 89.5 & 5.07 &&& 19000.0 & 23.9 & 2.30 \\
&& 23000.0 & 130.0 & 6.74 &&& 23000.0 & 63.5 & 6.03 \\
&& 27000.0 & 87.0 & 6.36 &&& 27000.0 & 28.9 & 4.39 \\
&& 31000.0 & 12.2 & 2.76 &&& 31000.0 & -9.5 & 1.02 \\\cline{3-5}  \cline{8-10}
 & 8.130 & 15000.0 & 43.5 & 2.92 && 8.13 & 15000.0 &  4.8 & 0.50 \\
&& 19000.0 & 116.1 & 5.90 &&& 19000.0 & 39.3 & 3.42 \\
&& 23000.0 & 164.3 & 7.28 &&& 23000.0 & 99.5 & 7.72 \\
&& 27000.0 & 120.2 & 7.00 &&& 27000.0 & 56.8 & 6.63 \\
&& 31000.0 & 30.4 & 4.58 &&& 31000.0 & -6.7 & 2.03 \\
\hline
\end{tabular*}}
\end{center}
\end{table}

\begin{table}
\centering{
\caption{ Results of Monte Carlo Simulations for N\,{\sc ii} at $ \log g$=4.0, and $\xi_t= 5.0\;\rm km\,s^{-1}$: average equivalent widths, $<W_\lambda>$, and the expected error, $2\,\sigma$
\label{mc_eqw_res}}}
\begin{center}
{\scriptsize
\setlength{\tabcolsep}{4pt}
\begin{tabular*}{0.51\textwidth}{l@{\hskip 0.09in}c@{\hskip 0.09in}c@{\hskip 0.04in}c@{\hskip  0.04in}c@{\hskip  0.09in}c@{\hskip  0.09in}c@{\hskip  0.09in}c@{\hskip  0.04in}c@{\hskip  0.04in}c}
\hline\hline
\multicolumn{1}{c}{$\lambda$}& $\epsilon_N$ & $\rm T_{eff}$ & \multicolumn{1}{l}{\tiny{$<W_{\lambda}>$}} & \multicolumn{1}{c}{$2\sigma$} &\multicolumn{1}{c}{$\lambda$}& $\epsilon_N$ & $\rm T_{eff}$ & \multicolumn{1}{l}{\tiny{$<W_{\lambda}>$}} & \multicolumn{1}{c}{$2\,\sigma$}\\     
\multicolumn{1}{l}{(\AA)}&& (K)& \multicolumn{1}{l}{(m\AA)} & (m\AA) &\multicolumn{1}{l}{(\AA)}&& (K)& \multicolumn{1}{l}{(m\AA)} & (m\AA) \\
\hline
3995 & 6.830 & 15000.0 &  3.0 & 0.32 & 6482 & 6.83 & 15000.0 &  0.1 & 0.01 \\
&& 19000.0 & 16.1 & 1.56 &&& 19000.0 &  1.3 & 0.16 \\
&& 23000.0 & 32.4 & 3.16 &&& 23000.0 &  5.3 & 0.74 \\
&& 27000.0 & 21.1 & 2.45 &&& 27000.0 &  2.8 & 0.57 \\
&& 31000.0 &  4.1 & 0.75 &&& 31000.0 & -1.3 & 0.15 \\\cline{3-5}  \cline{8-10}
 & 7.230 & 15000.0 &  6.6 & 0.66 && 7.23 & 15000.0 &  0.3 & 0.03 \\
&& 19000.0 & 30.9 & 2.53 &&& 19000.0 &  3.6 & 0.41 \\
&& 23000.0 & 60.6 & 4.68 &&& 23000.0 & 14.5 & 1.75 \\
&& 27000.0 & 44.7 & 4.29 &&& 27000.0 &  8.9 & 1.48 \\
&& 31000.0 & 12.7 & 1.87 &&& 31000.0 & -0.8 & 0.37 \\\cline{3-5}  \cline{8-10}
 & 7.530 & 15000.0 & 11.4 & 1.04 && 7.53 & 15000.0 &  0.6 & 0.07 \\
&& 19000.0 & 46.6 & 3.34 &&& 19000.0 &  7.1 & 0.76 \\
&& 23000.0 & 87.4 & 5.56 &&& 23000.0 & 28.0 & 2.98 \\
&& 27000.0 & 70.7 & 5.55 &&& 27000.0 & 19.8 & 2.79 \\
&& 31000.0 & 26.1 & 3.24 &&& 31000.0 &  1.4 & 0.86 \\\cline{3-5}  \cline{8-10}
 & 7.830 & 15000.0 & 18.5 & 1.53 && 7.83 & 15000.0 &  1.2 & 0.14 \\
&& 19000.0 & 66.1 & 4.19 &&& 19000.0 & 13.2 & 1.32 \\
&& 23000.0 & 116.9 & 6.28 &&& 23000.0 & 48.7 & 4.46 \\
&& 27000.0 & 101.4 & 6.42 &&& 27000.0 & 39.6 & 4.60 \\
&& 31000.0 & 47.2 & 4.78 &&& 31000.0 &  7.4 & 1.87 \\\cline{3-5}  \cline{8-10}
 & 8.130 & 15000.0 & 28.6 & 2.14 && 8.13 & 15000.0 &  2.4 & 0.26 \\
&& 19000.0 & 89.4 & 5.12 &&& 19000.0 & 23.0 & 2.11 \\
&& 23000.0 & 149.0 & 7.08 &&& 23000.0 & 76.6 & 5.98 \\
&& 27000.0 & 134.4 & 7.05 &&& 27000.0 & 69.5 & 6.54 \\
&& 31000.0 & 74.7 & 5.99 &&& 31000.0 & 20.7 & 3.54 \\
\hline
\end{tabular*}}
\end{center}
\vspace{0.1in}
\end{table}

\begin{table}
\centering{
\caption{ Results of Monte Carlo Simulations for N\,{\sc ii} at $\log g$=4.5, and $\xi_t= 5.0\;\rm km\,s^{-1}$: average equivalent widths, $<W_{\lambda}>$, and the expected error, $\sigma$
\label{mc_eqw_res_g4.5_xi_t5}}}
\begin{center}
{\scriptsize
\begin{tabular*}{0.51\textwidth}{l@{\hskip 0.09in}c@{\hskip 0.09in}c@{\hskip -0.05in}c@{\hskip  -0.05in}c@{\hskip  0.09in}c@{\hskip  0.09in}c@{\hskip  0.09in}c@{\hskip  -0.05in}c@{\hskip  -0.05in}c}
\hline\hline
\multicolumn{1}{c}{$\lambda$}& $\epsilon_N$ & $\rm T_{eff}$ & \multicolumn{1}{l}{\tiny{$<W_{\lambda}>$}} & \multicolumn{1}{c}{$2\sigma$} &\multicolumn{1}{c}{$\lambda$}& $\epsilon_N$ & $\rm T_{eff}$ & \multicolumn{1}{l}{\tiny{$<W_{\lambda}>$}} & \multicolumn{1}{c}{$2\,\sigma$}\\     
\multicolumn{1}{l}{(\AA)}&& (K)& \multicolumn{1}{l}{(m\AA)} & (m\AA) &\multicolumn{1}{l}{(\AA)}&& (K)& \multicolumn{1}{l}{(m\AA)} & (m\AA)\\
\hline
3995 & 6.830 & 15000.0 &  1.4 & 0.16 & 6482 & 6.83 & 15000.0 &  0.1 & 0.01 \\
&& 19000.0 &  9.3 & 0.95 &&& 19000.0 &  0.7 & 0.08 \\
&& 23000.0 & 24.4 & 2.40 &&& 23000.0 &  3.5 & 0.44 \\
&& 27000.0 & 23.2 & 2.53 &&& 27000.0 &  3.7 & 0.59 \\
&& 31000.0 &  8.4 & 1.13 &&& 31000.0 &  0.3 & 0.18 \\\cline{3-5}  \cline{8-10}
 & 7.230 & 15000.0 &  3.3 & 0.35 && 7.23 & 15000.0 &  0.1 & 0.02 \\
&& 19000.0 & 19.1 & 1.71 &&& 19000.0 &  1.8 & 0.21 \\
&& 23000.0 & 47.2 & 3.79 &&& 23000.0 &  9.5 & 1.08 \\
&& 27000.0 & 48.0 & 4.29 &&& 27000.0 & 11.0 & 1.47 \\
&& 31000.0 & 21.1 & 2.49 &&& 31000.0 &  2.4 & 0.54 \\\cline{3-5}  \cline{8-10}
 & 7.530 & 15000.0 &  6.1 & 0.60 && 7.53 & 15000.0 &  0.3 & 0.03 \\
&& 19000.0 & 30.6 & 2.43 &&& 19000.0 &  3.7 & 0.40 \\
&& 23000.0 & 70.1 & 4.76 &&& 23000.0 & 18.4 & 1.92 \\
&& 27000.0 & 74.4 & 5.46 &&& 27000.0 & 22.9 & 2.70 \\
&& 31000.0 & 38.4 & 3.89 &&& 31000.0 &  6.8 & 1.18 \\\cline{3-5}  \cline{8-10}
 & 7.830 & 15000.0 & 10.5 & 0.96 && 7.83 & 15000.0 &  0.6 & 0.07 \\
&& 19000.0 & 45.9 & 3.26 &&& 19000.0 &  7.1 & 0.74 \\
&& 23000.0 & 96.7 & 5.68 &&& 23000.0 & 32.8 & 3.04 \\
&& 27000.0 & 104.7 & 6.32 &&& 27000.0 & 42.9 & 4.31 \\
&& 31000.0 & 62.6 & 5.21 &&& 31000.0 & 16.2 & 2.31 \\\cline{3-5}  \cline{8-10}
 & 8.130 & 15000.0 & 17.3 & 1.44 && 8.13 & 15000.0 &  1.2 & 0.14 \\
&& 19000.0 & 65.2 & 4.22 &&& 19000.0 & 13.1 & 1.27 \\
&& 23000.0 & 126.4 & 6.76 &&& 23000.0 & 53.3 & 4.35 \\
&& 27000.0 & 137.2 & 7.13 &&& 27000.0 & 71.3 & 6.00 \\
&& 31000.0 & 91.5 & 6.17 &&& 31000.0 & 33.4 & 3.94 \\
\hline
\end{tabular*}}
\end{center}
\end{table}


\bsp	
\label{lastpage}
\end{document}